\documentclass[twoside]{IEEEtran}
\usepackage[table]{xcolor}
\usepackage{xspace,soul}
\usepackage{cite}
\ifCLASSINFOpdf
   \usepackage[pdftex]{graphicx}
\else
\fi
\usepackage[cmex10,]{amsmath}
\usepackage[export]{adjustbox}
\usepackage[caption=false,font=footnotesize]{subfig}

\usepackage[english]{babel}
\usepackage[utf8]{inputenc}
\usepackage[T1]{fontenc}

\usepackage[scaled=0.91]{helvet}
\usepackage[cmintegrals,varvw,]{newtxmath}
\usepackage{bm}
\usepackage{siunitx,booktabs}

\newcommand*\mat[1]{\textsf{\textit{\textbf#1}}}
\newcommand*\matelem[1]{\textsf{\textit{#1}}}
\renewcommand*\vec[1]{\textsf{\textit{\textbf#1}}}

\usepackage[nolist]{acronym}
\renewcommand*{\acsfont}[1]{\expandafter\MakeUppercase\expandafter{#1}}

\makeatletter
\renewcommand{\fnum@figure}{Fig. \thefigure}
\makeatother

\newcolumntype{L}[1]{>{\raggedright\arraybackslash}p{#1}}
\newcolumntype{C}[1]{>{\centering\arraybackslash}p{#1}}
\newcolumntype{R}[1]{>{\raggedleft\arraybackslash}p{#1}}

\begin{document}
\acrodef{AUT}[aut]{antenna under test}
\acrodef{DOF}{degrees of freedom}
\acrodef{MoM}[mom]{method of moments}
\acrodef{FIAFTA}[fiafta]{fast irregular antenna field transformation algorithm}
\acrodef{MLFMM}[mlfmm]{multi-level fast multipole method}
\acrodef{RWG}[rwg]{Rao-Wilton-Glisson}
\acrodef{MFIE}[mfie]{magnetic field integral equation}
\acrodef{EFIE}[efie]{electric field integral equation}
\acrodef{NFFFT}[NFFFT]{\ac{NF} to \ac{FF} transformation}
\acrodef{NF}{near-field}
\acrodef{FF}{far-field}
\acrodef{SVD}{singular-value decomposition}
\acrodef{CG}{conjugate gradient}
\acrodef{GMRES}{generalized minimum residual}
\acrodef{NE}{normal equation}
\acrodef{CS}{combined-source}
\acrodef{PEC}{perfectly electrically conducting}
\acrodef{SNR}{signal-to-noise ratio}
\acrodef{EVD}{eigenvalue decomposition}
\acrodef{MVP}{matrix-vector product}
\acrodef{CP}{Calderón projector}
\acrodef{WFCP}[WF-CP]{weak-form \ac{CP}}
\acrodef{NRE}{normal residual system of equations}
\acrodef{NEE}{normal error system of equations}
\acrodef{OEWG}{open-ended waveguide}
\acrodef{SV}{singular value}
\acrodef{OE}{observation error}
\acrodef{RD}{reconstruction deviation}
\acrodef{SH}{spherical harmonics}
\acrodef{SC}{side constraint}
\acrodef{D-SH}{distributed \ac{SH}}

\title{Accuracy and Conditioning of Surface-Source Based  Near-Field to Far-Field Transformations}

\author{\IEEEauthorblockN{
Jonas Kornprobst, \IEEEmembership{Graduate Student Member, IEEE}, 
Josef Knapp, \IEEEmembership{Graduate Student Member, IEEE}, 
Raimund A. M. Mauermayer, \IEEEmembership{Graduate Student Member, IEEE},
Ole Neitz, \IEEEmembership{Graduate Student Member, IEEE}, 
Alexander Paulus, \IEEEmembership{Graduate Student Member, IEEE},  
and Thomas F. Eibert, \IEEEmembership{Senior Member, IEEE}
}%
\thanks{Manuscript received April 21, 2020; revised September 28, 2020; accepted December 05, 2020; date of current version January 15, 2021.}
\thanks{The authors are with the Chair of High-Frequency Engineering, Department of Electrical and Computer Engineering, Technical University  of  Munich,  80290  Munich, Germany (e-mail:  j.kornprobst@tum.de;hft@ei.tum.de).}%
\thanks{Color versions of one or more of the figures in this paper are available online at http://ieeexplore.ieee.org}%
\thanks{Digital Object Identifier 10.1109/TAP.2020.3048497}
}

\maketitle

\begin{abstract}
The conditioning and accuracy of various inverse surface-source formulations are investigated. 
First, the normal systems of equations are discussed. 
Second, different implementations of the zero-field condition are analyzed regarding their effect on  solution accuracy, conditioning, and source ambiguity.  
The weighting of the Love-current side constraint is investigated in order to provide an accurate problem-independent methodology. 

The transformation results for simulated and measured near-field data show a comparable behavior regarding accuracy and conditioning for most of the formulations. 
Advantages of the Love-current solutions are found only in diagnostic capabilities. 
Regardless of this, the Love side constraint is a computationally costly way to influence the iterative solver threshold, which is more conveniently controlled with the appropriate type of normal equation.

The solution behavior of the inverse surface-source formulations is mostly influenced by the choice of the reconstruction surface. A spherical Huygens surface leads to the best conditioning, whereas the most accurate solutions are found with a tight, possibly convex hull around the antenna under test.
\end{abstract}


\section{Introduction} %
\IEEEPARstart{I}{n general}, the antenna radiation properties of interest are \ac{FF} quantities, such as gain or radiation pattern. 
In order to retrieve the \ac{FF} properties, \ac{NF} measurements offer a versatile method. 
As such, the measurement values are used to solve an inverse-source problem first and all desired quantities are then calculated from the reconstructed sources. 
Such an \aclu{NFFFT} (\acs{NFFFT}) can provide a so-called \emph{processing gain} with respect to the unavoidable measurement errors, 
which would directly affect \ac{FF} measurement results. 
In source reconstruction methods, statistical errors are averaged and filtered if a suitable reconstruction basis is used which only provides a limited number of \aclu{DOF} (DOFs) for the radiated fields.  
With more measurement samples than dictated by the sampling limit, more robustness against measurement noise and other errors is attained in \ac{NFFFT}s.

Equivalent surface-current models, discretized by a surface mesh on a Huygens surface enclosing the \ac{AUT}, give the possibility to incorporate geometrical information about the \ac{AUT} and provide diagnostic and spatial filtering capabilities \cite{Qui2010,Jorg2011,Foged2014,Parini2014,Eibert2016,Kornprobst2019Love}. 
The equivalent source representation with electric and magnetic surface current densities is not unique~\cite{Alvarez2007}, as we know from the equivalence theorems~\cite{Huygens1690,Love1901, Schelkunoff1936,4555260}. 
Therefore, the inverse problem is ill-posed. 
We can choose purely electric or magnetic surface current densities~\cite{Petre1992,Petre1994,Qui2010,Quijano2010} or  
combinations of electric and magnetic surface current densities such as Love currents~\cite{AraqueQ2009,Jorg2010,Qui2010,Quijano2010,Jorg2011,Jorg2012,Foged2014,Kilic2015,Kornprobst2019} or combined sources (\acsu{CS}s)~\cite{Eibert2016}. 

In literature, the Love condition is sometimes claimed to exhibit a better conditioning and a more stable and accurate solution behavior than other equivalent current methods~\cite{AraqueQ2009,Qui2010,Quijano2010,Jorg2011a}. 
The improved solution stability  might hold true dependent on the solver and its implicit regularization properties~\cite{Qui2010}. Unconstrained current solutions may vary arbitrarily as compared to Love currents and may, thus, depend strongly on measurement errors.
However, according to the equivalence principle, the external fields (both near and far) of all equivalent current scenarios are indistinguishable.
Hence, the solution stability of the retrieved currents is not very meaningful for the field reconstruction problem 
as long as the current variations do not cause additional numerical errors in the solution process, e.g., by numerical cancellation.

In this paper, we describe the inverse surface-source problem of antenna \ac{NF} measurements and its fusion with four different implementations of the Love condition: the \ac{CS} constraint, a post-processing technique, a \ac{SC}, and the \ac{CP}. 
Parts of these formulations have been covered in~\cite{Kornprobst2019,Kornprobst2019Love}. 
In this work, we analyze  the reconstruction accuracy and the zero-field quality, i.e., the accuracy of the Love condition, more rigorously. This includes the discretization of both the \ac{SC} and the \ac{CP}, and the accuracy implications on the retrieved fields and currents.
This issue is analyzed by \ac{SV} spectra and detailed iterative solver convergence analyses.

The paper is structured as follows. In Section~\ref{sec:general}, the integral equations are introduced in continuous and discretized form. Section~\ref{sec:fiafta} gives an overview of the \ac{FIAFTA} with a focus on equivalent current reconstruction. 
The two different systems of normal equations (\acsu{NE}s) are discussed to highlight the different kinds of $\ell^2$-regularized iterative solutions of the inverse problem. 
In Section~\ref{sec:ZF}, the \ac{CP} and the Love current \ac{SC} are discretized as linear systems of equations providing several variants of the zero-field enforcement. The \ac{CS} condition, an approximation of the Love condition under certain circumstances, and a Love-current post-processing technique are briefly revisited. 
Section~\ref{sec:scale} discusses the weighting factor for the Love \ac{SC}. A methodology dependent on the \ac{OE} is proposed. 
Finally, the accuracy and conditioning of surface-source based \ac{NFFFT}s are investigated for both simulation data with known reference  solution and \ac{NF} measurement data without the knowledge of the true solution.

\section{Boundary Integral Equation Notation\label{sec:general}} 
\begin{figure}[t]
\centering
\includegraphics{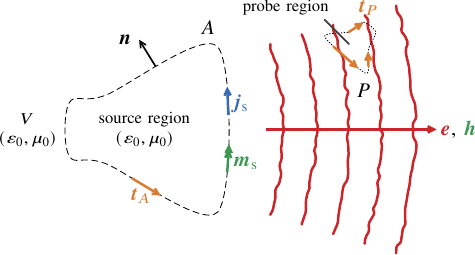}
\caption{The considered general equivalent source scenario with equivalent electric and magnetic surface current densities $\bm j_\mathrm s$ and $\bm m_\mathrm s$ on the Huygens surface $A$. The surface testing functions are either defined as $\bm t_A$  directly on the AUT surface $A$  or as $\bm t_P$ on a surface around a probe somewhere in $V$.\label{fig:problem2}\label{fig:problem}}
\end{figure}

Consider a boundary value radiation problem as shown in Fig.~\ref{fig:problem}. The radiation originates from the source region enclosing the \ac{AUT} and is described by equivalent sources on its surface $A$ (or inside this region) according to the Huygens principle~\cite{Huygens1690, Love1901, Schelkunoff1936}. 
Prescribed boundary values may be defined anywhere on $A$ (e.g., for a scattering problem) or outside of $A$. 
For \ac{NF} antenna measurements, boundary values are defined in the form of measurement values of the probe antenna measurement signals. 
They are evaluated by weighting functions on the probe hull surfaces $P$.
For the zero-field condition, the radiated fields are evaluated on $A$.
The time-harmonic case with suppressed $\mathrm e^{\,\mathrm j \omega t}$ time convention is assumed throughout the paper.

A word in front:
This paper employs several uncommon but important abbreviations related to the inverse problem formulation and the chosen type of sources. All these abbreviations are summarized in the appendix in Tab.~\ref{tab:acs}.

\subsection{Continuous Description of the Equivalent Surface Sources}
To reconstruct the radiated fields, equivalent electric and magnetic surface current densities $\bm j_\mathrm s$ and $\bm m_\mathrm s$ are defined on the Huygens surface $A$ (with the outward unit normal~$\bm n$) surrounding the source region. 
To avoid conditioning issues, $Z_0\bm j_\mathrm s$ are employed as unknowns instead of $\bm j_\mathrm s$, i.e., the electric currents are scaled with the free-space wave impedance $Z_0$. 
Then, all surface current densities exhibit the same unit of magnetic surface current densities. 
This results in the integral representation of the electric field
\begin{equation}
\bm e(\bm r) =   \iint\nolimits_A\!\Big[ Z_0^{{-1}}
\bm{\mathcal {G}} ^{\,\bm e}_{\bm j}(\bm r , \bm r') \cdot Z_0\bm { j}_\mathrm s (\bm r')+
\bm{\mathcal {G}} ^{\,\bm e}_{\bm m}(\bm r , \bm r') \cdot \bm { m}_\mathrm s (\bm r')
\Big]
\mathop{}\!
\mathrm{d}a'\,, \label{eq:efield}
\end{equation}
where $\bm{\mathcal {G}} ^{\,\bm e}_{\bm j}$ and $\bm{\mathcal {G}} ^{\,\bm e}_{\bm m}$ are the dyadic Green's functions for a source at $\bm r'$ and observation at $\bm r$. 
Employing duality, we find a similar relationship 
\begin{equation}
Z_0\bm h (\bm r)= \iint\nolimits_A\!\Big[ Z_0 \,
\bm{\mathcal {G}} ^{\,\bm h}_{\bm m}(\bm r , \bm r') \cdot \bm { m}_\mathrm s (\bm r')+
\bm{\mathcal {G}} ^{\,\bm h}_{\bm j}(\bm r , \bm r') \cdot Z_0\bm { j}_\mathrm s (\bm r')
\Big]
\mathop{}\!\mathrm{d}a'\label{eq:magneticfield}
\end{equation}
for the magnetic field, which also receives a scaling with $Z_0$ to avoid conditioning issues. Both field types have the dimension of an electric field. 
For the sake of a more compact notation, we introduce field calculation integral operators for a surface current density $\bm \beta$, which can represent either $\bm m _\mathrm s$ or $Z_0\bm j_\mathrm s$,
\begin{IEEEeqnarray}{rl}
\bm{\mathcal{T}}(\bm r)\{\bm { \beta}\} \coloneq&~
Z_0^{{-1}}\bm n \times\iint_A
\bm{\mathcal {G}} ^{\,\bm e}_{\bm j}(\bm r , \bm r') \cdot \bm { \beta} (\bm r')
\mathop{}\!\mathrm{d}a'\notag\\
=&~
{Z_0^{\phantom{-1}}}\bm n \times\iint_A 
\bm{\mathcal {G}} ^{\,\bm h}_{\bm m}(\bm r , \bm r') \cdot \bm { \beta} (\bm r')
\mathop{}\!\mathrm{d}a'\,,
\end{IEEEeqnarray}
\begin{IEEEeqnarray}{rl}
\bm{\mathcal{K}}(\bm r)\{\bm { \beta}\}  \coloneq& ~
{-}\bm n\times\iint_A
\bm{\mathcal {G}} ^{\,\bm h}_{\bm j}(\bm r , \bm r') \cdot \bm { \beta} (\bm r')
\mathop{}\!\mathrm{d}a'
\notag\\
=&~\phantom{-}\bm n\times\iint_A 
\bm{\mathcal {G}} ^{\,\bm e}_{\bm m}(\bm r , \bm r') \cdot \bm { \beta} (\bm r')
\mathop{}\!\mathrm{d}a'\,.
\end{IEEEeqnarray}
The evaluation of the tangential fields according to~\eqref{eq:efield} and~\eqref{eq:magneticfield} for observation points $\bm r$ on the surface $A$, thus, yields
\begin{equation}
\bm n \times\bm e(\bm r) =   
\bm{\mathcal {T}} (\bm r)\{Z_0\bm { j}_\mathrm s (\bm r')\}+
\begin{bmatrix} -\frac{1}{2}\bm{\mathcal {I}}(\bm r) + \bm{\mathcal {K}}(\bm r) \end{bmatrix} \bm { m}_\mathrm s (\bm r')
\,,
\end{equation}
\begin{equation}
Z_0\bm n \times\bm h (\bm r)= 
\bm{\mathcal {T}}(\bm r)\{\bm { m}_\mathrm s (\bm r')\}+
\begin{bmatrix} \frac{1}{2}\bm{\mathcal {I}}(\bm r) - \bm{\mathcal {K}}(\bm r) \end{bmatrix} Z_0\bm { j}_\mathrm s (\bm r')
\,,
\end{equation}
where the identity operators
\begin{IEEEeqnarray}{rl}
\bm{\mathcal{I}}(\bm r)\{\bm { \beta}\}  \coloneq \bm \beta(\bm r)
\end{IEEEeqnarray}
result from the integration of the singularity of the Green's function in the $\bm{\mathcal K}$-operator, which is now evaluated in a Cauchy principal value sense.

\subsection{General Discretization Strategy}
With a triangular mesh on $A$, the surface current densities are modeled by \ac{RWG} functions as~\cite{Rao1982}
\begin{equation}
Z_0\bm j_\mathrm s = \sum\nolimits_{n=1}^N\bm\beta_n[Z_0 \vec i\hspace*{0.05cm}]_n 
\,,\qquad
\bm m_\mathrm s = \sum\nolimits_{n=1}^N\bm\beta_n[\vec v\hspace*{0.01cm}]_n 
\label{eq:jm}
\end{equation}
on pairs of neighboring triangles. 
The electric current unknowns  $\vec i$ and the magnetic ones $\vec v$ are, if used at the same time, concatenated to the unkowns vector $\vec x = [Z_0 \vec i ~\vec v]^\mathrm T$. 
In a rather general way, the field evaluation is then performed by surface density testing functions $\bm t(\bm r)$, see Fig.~\ref{fig:problem2}.
Evaluating $\bm n\times \bm e$ radiated by an electric surface current density basis function (or $Z_0\bm n \times \bm h$ for a magnetic current basis function) with one specific testing function $\bm t_m$, we obtain one element of a forward operator matrix  as
\begin{equation}
[\mat T_{\bm t}]_{mn}
=
\iint_{A,P} \bm t_m (\bm r) \,\cdot \bm{\mathcal{T}}(\bm r)\{\bm { \beta}_n\}
\mathop{}\!\mathrm{d}a\,.
\end{equation}
The same procedure for the magnetic field $Z_0\bm h$ of an electric surface current basis function, and the electric field $\bm e$ of a magnetic surface current basis function, leads to

\begin{equation}
{[}\mat K_{\bm t}]_{mn}
={-}\iint_{A,P} \bm t_m (\bm r) \,\cdot \bm{\mathcal{K}}(\bm r)\{\bm { \beta}_n\}
\mathop{}\!\mathrm{d}a\,.
\label{eq:mfie}
\end{equation}
The $\bm{\mathcal I}$ operator is represented by a Gram matrix with the entries
\begin{equation}
[\mat G_{\hspace*{0.04cm}\bm {t\beta}}]_{mn}\,\,
=\iint_A \bm t_m (\bm r) \,\cdot \bm{\mathcal I}(\bm r)\{ \bm { \beta}_n (\bm r)\}
\mathop{}\!\mathrm{d}a\,.
\end{equation}
Testing the electric field gives a a right-hand side vector of the equation system with the entries
\begin{equation}
[\vec b]_{m}=
\iint_{A,P} \bm t_m (\bm r) \cdot
\big (
\bm n \times \bm e (\bm r) 
\big)
\mathop{}\!\mathrm{d}a+[\bm\epsilon_\textsc{oe}]_m\,,\label{eq:transmission}
\end{equation} 
where\,---\,for the case of measurements\,---\,the vector $\bm\epsilon_\textsc{oe}$ includes the \ac{OE} or measurement error arising, e.g., from noise or positioning imperfections.

\section{Equivalent Source Reconstruction\label{sec:fiafta}} 
\subsection{Description of the Forward Operator}
For the source reconstruction, the received power wave at the probe is evaluated in the presented formulation by testing the radiated fields  with an equivalent current representation of the probe (or, equivalently, with  a spectral representation in the accelerated implementation of the \ac{FIAFTA}).
We have full freedom to choose any current representation of the probe.
Then, the discretized inverse problem reads
\begin{equation}
\mat A \,\vec x =
\begin{bmatrix}
\mat T_{\bm t_P}& \mat K_{\bm t_p}
\end{bmatrix} \vec x =
\vec b
\,,\label{eq:fiafta}
\end{equation}
where $\bm t_m = \bm t_{P\!,m}$ is chosen to be an equivalent electric surface current description of the probe antenna employed for the $m$th measurement, correctly rotated and shifted in space to the measurement location $\bm r_m$. 

For the equivalent current description, the \ac{AUT} geometry is enclosed by the Huygens surface exactly as depicted in Fig.~\ref{fig:problem}. 
According to the equivalence principle, we can choose a source description with $\bm j_\mathrm s$ and $\bm m_\mathrm s$ or alternatively one with $\bm j_\mathrm s$  only, dropping the $\bm m_\mathrm s$  unknowns (or vice versa). 

Assuming we know a true solution~$\hat{\hspace*{-0.035cm}\vec x}$ of the inverse problem, imperfect measurements cause the \ac{OE} in the \ac{NF}
\begin{equation}
\epsilon_\textsc{oe}=
\frac{\lVert\mat A\, \hat{\hspace*{-0.035cm}\vec x} - \vec b\rVert_2}{\lVert\vec b\rVert_2}=
\frac{\lVert\bm \epsilon_\textsc{oe}\rVert_2}{\lVert\vec b\rVert_2}\,.
\end{equation}
Some part of this error may be reconstructed by a false solution contribution, which is linearly superimposed to the correct solution. 
The remainder of the \ac{OE} is not attributable to any~$\vec x$. 
Solving~\eqref{eq:fiafta} retrieves thus a solution~$\vec x$, but at the observation locations, there remains a \ac{RD} 
\begin{equation}
\epsilon_\textsc{rd}=
\frac{\lVert\mat A\, \vec x - \vec b\rVert_2}{\lVert\vec b\rVert_2}\,,
\end{equation}
which estimates the \ac{OE} and helps to suppress errors. 

The number of unknowns $N_\mathrm{un}$ equals either $N$ or $2N$ and defines the length of the vector $\vec x$ and the number of columns of $\mat A$; the number of observation points $N_\mathrm{ms}$, which is the number of rows in $\mat A$ and at the same time the number of entries in $\vec b$, does not match  to $N_\mathrm{un}$ in general, but  the matrix $\mat A$ is wide for common antenna \ac{NF} measurement setups with subsequent equivalent current \ac{NFFFT}s, i.e., $N_\mathrm{ms} < N_\mathrm{un}$. 
This is due to the fact that the number of measurements necessary for a correct \ac{NF} reconstruction $N_\mathrm{ms}$ is chosen according to the number of \ac{DOF}s $N_\mathrm{dof}$ of the radiation fields which in turn solely depend on the size and shape of the \ac{AUT}. 
E.g., if a field representation with spherical modes is chosen, the number of radiating modes (i.e.,  the possible \ac{DOF}s in the fields) is limited by the minimum sphere around the \ac{AUT}~\cite{Hansen1988}. 
Often, $N_\mathrm{ms}$ is chosen a bit larger than $N_\mathrm{dof}$ to avoid  aliasing in the \ac{AUT} mode spectrum and to cope with measurement errors. 

This certainly leads to a nontrivial cokernel (also called left nullspace) $\mathrm{ker}\,\mat A ^\mathrm H$, since $\mathrm{rank}\,\mat A \approx N_\mathrm{dof}$.
We find the reconstructed solution in the image $\mathrm{im}\,\mat A$. Any part of $\vec b$ in  $\mathrm{ker}\,\mat A ^\mathrm H$ cannot be reconstructed and leads to a RD. 
This is desired, since the RD should contain all measurement errors and noise, i.e., contributions which cannot be mapped onto the equivalent sources and which are thus suppressed. 
With $N_\mathrm{ms} > N_\mathrm{dof}$, we have  a non-vanishing cokernel dimension and the system of equations in~\eqref{eq:fiafta} can be considered to be overdetermined despite $\mat A$ being wide.

In addition to the oversampling of observations, the inverse equivalent surface source problem exhibits another, totally unrelated, kind of oversampling in the unknowns space. 
The equivalent currents are modeled on a (possibly non-convex) mesh where $\lambda/10$ is a typical discretization density.
This source representation is by far oversampled due to a large $N_\mathrm{un}$ as compared to the \ac{DOF}s of the \ac{AUT}, which can be roughly approximated by $\lambda/2$-spaced measurements on the minimum sphere or a smaller AUT hull. 
A similar oversampling is present in a \ac{D-SH} expansion. 
Due to this oversampling, evanescent modes are possibly excited. 
However, they are impossible to reconstruct even with infinitely precise algorithms since they are typically not observable at the measurement distance due to uncertainties. 
In a \ac{MLFMM}-accelerated forward operator, strongly evanescent modes are not propagated to the measurement locations due to the inherent low-pass filtering.
All of this implies a non-trivial null space $\mathrm {ker}\,\mat A$, i.e., the solution of~\eqref{eq:fiafta} cannot be unique and  the inverse problem is thus mildly ill-posed.

Additionally, various surface current representations (purely electric or magnetic, both, etc.) exist which obviously differ by non-radiating currents, i.e., they differ by solutions of the corresponding interior problem. Changing the retrieved solution by non-radiating currents has no effect on any fields outside of $A$ for reasonably chosen shapes of $V$; neither on radiating nor on evanescent modes. 
Such an inverse problem can be seen as severely ill-posed.

Both effects cause a non-trivial $\mathrm {ker}\,\mat A$ (with a larger dimension for ambiguous electric and magnetic currents). Unknown vectors with an effect on the reconstruction are only found in $\mathrm{im}\,\mat A^\mathrm H$. In the sense of $N_\mathrm{un}\gg N_\mathrm{dof}$, 
we conclude that the equation system~\eqref{eq:fiafta} is underdetermined.
Typically, regularization is employed to get rid of negative effects of $\mathrm {ker}\,\mat A$.

The unknowns ambiguity is commonly removed by minimizing a certain norm of the currents or \ac{NF} residuals. 
A common way is the solution of~\eqref{eq:fiafta} in a least-squares sense by minimizing an $\ell^2$-norm, which is discussed in the next section. 
Only in special scenarios, other norms might be of interest~\cite{Hofmann2019}. 
In summary, we find that the band-limited forward operator of inverse surface-source problems with oversampled measurements exhibits a non-trivial kernel and a non-trivial cokernel. Both have to be kept in mind to appropriately solve the inverse surface-source problem. Due to the typical surface-source oversampling and the existence of non-radiating currents, we commonly have $\mathrm {dim}(\mathrm {ker}\,\mat A)\gg\mathrm {dim}(\mathrm {ker}\,\mat A^\mathrm H)$.

\subsection{The Normal Equations}
To solve~\eqref{eq:fiafta}, direct methods are not suitable due to their high complexity and incompatibility with fast algorithms.
To reduce the computational complexity, iterative solvers are the method of choice in conjunction with well-conditioned fast formulations to obtain a solution with $\mathcal O(N_\mathrm{it}N_\mathrm{un}\log N_\mathrm{un})$ complexity, where $N_\mathrm{it}\ll N_\mathrm{un}$. 
Since standard iterative solvers such as the \ac{CG} method~\cite{hestenes1952methods} or the \ac{GMRES} method~\cite{saad1986gmres} have been initially proposed for square matrices, which are commonly not encountered for measurement scenarios, 
\ac{NE}s are the method of choice to resolve this issue and to obtain a square system matrix~\cite{Saad2003}. 
The common formulation employed in antenna \ac{NF} measurements is
\begin{equation}
\mat A ^\mathrm H \mat A \,\vec x =\mat A ^\mathrm H \vec b \,,\label{eq:normal1}
\end{equation}
where the adjoint of the forward operator $\mat A ^\mathrm H$ is multiplied from the left-hand side. 
Eq.~\eqref{eq:normal1} is called a \ac{NRE} and suitable for overdetermined systems of equations~\cite{Saad2003}. 
The overdeterminedness of the inverse problem is resolved by minimizing the $\ell^2$-norm of the residual of \eqref{eq:fiafta}, i.e., by minimizing the \ac{RD} $\rVert\mat A\,\vec x - \vec b\lVert_2$~\cite{Saad2003}. 
Hence, the \ac{NRE} takes care of a possibly non-trivial $\mathrm{ker}\,\mat A^\mathrm{H}$. 
The problem of a non-trivial $\mathrm{ker}\,\mat A$ for underdetermined problems persists and additional regularization is necessary for a stable solution.
The employed solver may impose additional regularization constraints to remove any ambiguity in the solution. In the case of \ac{GMRES}, we observe an $\lVert\vec x\rVert_2$ regularization with a suitable termination threshold~\cite{calvetti2002regularizing,elden2012gmres}.
A truncated \ac{SVD} has a similar effect.

Even though the \ac{NRE} is very common, the typical equivalent-current \ac{NFFFT} scenario resembles more an underdetermined system than an overdetermined system, see the discussion in the previous subsection.
Accordingly, the better suited \ac{NEE} reads~\cite{Saad2003}
\begin{equation}
\mat A\, \mat A ^\mathrm H \,\vec y = \vec b \,,\label{eq:normal2}
\end{equation}
where the adjoint operator is multiplied prior to the standard forward operator. 
To retrieve the solution of the inverse problem, the post-processing step
\begin{equation}
\vec x = \mat A ^\mathrm H \,\vec y  \label{eq:normal3}
\end{equation}
is necessary. 
Several important differences are observed. 
First, the iterative solution works with the vector $\vec y$, an auxiliary NF vector at the observation locations.
Second, the \ac{NF} RD $\ell^2$-norm is evaluated during the solution process\,---\,and not the $\ell^2$-norm of the current residual. 
This is advantageous since no arbitrary termination criterion for the current residual has to be defined.  
In contrast, the iterative solution stops if the \ac{NF} error approaches the possible minimum, limited by the \ac{OE}. 
From a theoretical point of view, this kind of \ac{NE} finds a different least-squares solution since it takes care of a non-trivial $\mathrm {ker}\,\mat A$: It minimizes the $\ell^2$-error norm of the unknowns vector, i.e., the norm $\rVert\mat A^\mathrm{H}\,\vec y - \hat{\hspace*{-0.035cm}\vec x}\lVert_2$ for  any correct solution $\hat{\hspace*{-0.035cm}\vec x }$ of \eqref{eq:fiafta}.
Additional regularization of $\vec y$ may also be employed; however, this does not influence the solution except for negligible numerical effects since any solution part in the column null-space of $\mat A$ is suppressed in~\eqref{eq:normal3}.

Both \ac{NE}s give similar solutions since the difference only consists of current components which cause no difference in the fields at the observation locations. 
However, the NEE offers advantages in the iterative solution process~\cite{kornprobst_measurementerror_2019}.

\section{Zero-Field Enforcement\label{sec:ZF}} 
The equivalent source description on any closed  Huygens surface is unique (unambigous) if one type of currents is utilized, i.e., either electric or magnetic surface current densities. 
Choosing both electric and magnetic surface current densities, additional constraints can be enforced to restrict the solution space and arrive at mildly ill-posed problem. 
One possible constraint is the Love condition, where the surface current densities 
\begin{equation}
\bm j_\mathrm{L} =\bm n \times \bm h\,,\qquad  \bm m_\mathrm{L}=\bm e \times \bm n 
\label{eq:love_field}
\end{equation}
represent the total tangential fields on the Huygens surface. 
This Love-current representation produces zero fields inside the source region, i.e., the \ac{AUT} volume. 
All other surface current solutions result from a superposition of non-radiating currents with interior fields only. 
In the following, four (two approximate and two exact) possibilities for the zero-field enforcement are described. 
\subsection{Combined-Source Approximation}
An approximation of the Love condition is given by the \ac{CS} condition, enforcing locally outward-oriented radiation~\cite{Mautz1979,Eibert2016}. 
This works if the surface under consideration contains all sources of the scenario (for more general scenarios with an impinging field, it is not a zero-field approximation any more~\cite{Kornprobst2018a}) and if the surface is convex and sufficiently smooth. 

We briefly discuss the \ac{CS} equivalent surface current representation since it is employed for comparison to the other methods. 
Assuming general equivalent surface currents according to~\eqref{eq:jm} (ignoring~\eqref{eq:love_field}),  the magnetic surface  current densities are obtained via
\begin{equation}
\bm m_\mathrm{cs} = Z_0 \,\bm n \times \bm j_\mathrm{cs} \label{eq:cs}
\end{equation}
from the electric surface current densities.
Due to the directive, outward-oriented radiation characteristic of the resulting sources, a null field inside the source region is approximated for convex Huygens surfaces.

The discretization of~\eqref{eq:cs} is achieved either by enforcing the $\bm n \times$-rotation of one type of surface current density or by a mapping between two sets of the same basis functions, e.g., RWG~\cite{Eibert2016,Eibert2017,Kornprobst2018a}. 
We follow the latter strategy since it was demonstrated to be more accurate for inverse equivalent surface-source scenarios~\cite{Eibert2017}.

\subsection{Love-Current Retrieval via Post-Processing}
After retrieving arbitrary equivalent surface currents (e.g., unconstrained electric and magnetic surface currents), the fields are calculated slightly outside of $A$ for each triangle and mapped back on the basis functions for $\bm j$ and $\bm m$. 
Hence, the inverse problem is solved and~\eqref{eq:love_field} is fulfilled. Alternatively, the Love-current mapping in subsection IV.\emph{C} can be performed just after we have obtained a solution~\cite{Kornprobst2019Love}.
\subsection{Love-Current Mapping via Calderón Projection}
It is possible to obtain Love currents by evaluating the tangential electric and magnetic fields on the Huygens surface, which in turn relate to the Love surface current densities $\bm j _\mathrm L$ and $\bm m _\mathrm L$. 
This  is known as \ac{CP}, written in the form of~\cite{Calderon1963,Hsiao1997,Nedelec2001,Kornprobst2019}
\begin{equation}
\begin{bmatrix}Z_0\bm j_\mathrm{L} \\ \bm m_\mathrm{L}\end{bmatrix}
=
\begin{bmatrix}Z_0\bm n \times \bm h \\ \bm e \times \bm n \end{bmatrix}
=
\begin{bmatrix} \frac{1}{2}\bm{\mathcal I} -  \bm{\mathcal K}  & \bm{\mathcal{T}}\\ 
                           - \bm{\mathcal{T}}                   & \frac{1}{2}\bm{\mathcal I} -  \bm{\mathcal K}  \end{bmatrix}
\begin{bmatrix} Z_0\bm j_\mathrm{s} \\ \bm m_\mathrm{s} \end{bmatrix}
\label{eq:LoveCP}
\end{equation}
and discretized as a mapping matrix
\begin{multline}
\hspace*{-0.35cm}\mat G\, \vec x_\mathrm L
=
\begin{bmatrix}\mat G_{\bm\beta\bm\beta} & \mat 0 \\ \mat 0 & \!\mat G_{\bm\beta\bm\beta} \end{bmatrix}
\begin{bmatrix}Z_0\vec i_\mathrm L \\ \vec v_\mathrm L\end{bmatrix}
=\\=
\begin{bmatrix} \frac{1}{2}\mat G_{\bm\beta\bm\beta}  +  \mat K_{\bm\beta}  & \mat T_{\bm\beta}\\ 
                            -\mat T_{\bm\beta} & \frac{1}{2}\mat G_{\bm\beta\bm\beta}  +  \mat K_{\bm\beta}
\end{bmatrix}
\begin{bmatrix} Z_0\vec i \\ \vec v\end{bmatrix}
= \mat L_\mathrm m \,\vec x
\label{eq:LoveCP_discr}
\end{multline}%
 with $\bm\beta$ testing functions for the rotated fields $\bm e \times \bm n$ and $\bm n \times \bm h$.
The diagonal blocks of  $\mat L_\mathrm m$ are a  well-conditioned \ac{MFIE} alike matrix.
Even if RWGs might not be the optimal choice of testing functions, this kind of testing is necessary since the mapping from the evaluated fields back to the currents is only possible due to the Gram matrices on the left-hand side of~\eqref{eq:LoveCP_discr}.\!\footnote{Buffa-Christansen testing functions are also feasible, $\bm n \times$RWGs are not.} 
The Gram matrix $ \mat G_{\bm\beta\bm\beta}$ on the left side of~\eqref{eq:LoveCP_discr} is easily inverted iteratively. 
The conditioning of this formulation is excellent due to the Gram matrices. 

However, it is clear that this \ac{MFIE}-alike discretization of the mapping operator may suffer from the discretization inaccuracies of the standard \ac{MFIE}~\cite{Chen1990,Cools_2011,Kornprobst2018b}. 
We investigate the accuracy improvement attained in an improved discretization of the identity operator~\cite{Kornprobst2018b}. In this case, the weak-form discretization of the identity operator inside the surface field evaluation reads
\begin{equation}
\mat G_{\bm\beta\bm\beta,\mathrm{wf}} = \frac{1}{2}\mat G_{\bm\beta\bm\beta}^{\phantom{-1}} - \frac{1}{2}\mat G_{\bm\beta\bm\alpha}^{\phantom{-1}}\mat G_{\bm\beta\bm\beta}^{-1}\mat G_{\bm\beta\bm\alpha}^{\phantom{-1}} \,,
\end{equation}
\begin{equation}
\mat G_{\mathrm{wf}} = 
\begin{bmatrix}\mat G_{\bm\beta\bm\beta,\mathrm{wf}} & \mat 0 \\ \mat 0 & \mat G_{\bm\beta\bm\beta,\mathrm{wf}}
\end{bmatrix}\,
\end{equation}
which changes the \ac{WFCP} matrix to
\begin{equation}
 \mat L_{\mathrm m,\textsc{WF-CP}}
 =
\begin{bmatrix} \frac{1}{2}\mat G_\mathrm{wf}  +  \mat K_{\bm\beta} & \mat T_{\bm\beta}\\ 
                            -\mat T_{\bm\beta} & \frac{1}{2}\mat G_\mathrm{wf}  +  \mat K_{\bm\beta}
\end{bmatrix}\,.
\end{equation}

Both types of the \ac{CP} are  applied to the \ac{NRE}~\eqref{eq:normal1} in the form of a left preconditioner as
\begin{equation}
\mat G^{-1} \,\mat L_\mathrm m\,\mat A ^\mathrm H (\mat A \,\vec x - \vec b)=\vec 0 \,.
\label{eq:normal1_Love1}
\end{equation}
The \ac{NRE} maps the field residual ($\mat A \,\vec x_i - \vec b$) of the $i$th solver iteration back to the current unknowns by the adjoint operator~$\mat A ^\mathrm H $.  
The subsequent \ac{CP} has the effect that the residual 
\begin{equation}
\vec r_i=\mat G^{-1} \,\mat L_\mathrm m\,\mat A ^\mathrm H (\mat A \,\vec x_i - \vec b)
\end{equation}
of the $i$th search vector $\vec x_i$\,---\,and, thus, also the final solution\,---\, 
 mostly contain Love currents with zero field inside the source region, where the inner-field suppression is limited by the current discretization and the accuracy of the \ac{CP}. 

An important effect of the \ac{CP} is that the ambiguity of choosing both electric and magnetic equivalent currents is eliminated. 
Possible benefits are analyzed in the results section.
For the \ac{NEE}, the very same mapping is introduced as
\begin{equation}
\mat A\,\mat G^{-1} \,\mat L_\mathrm m\,\mat A ^\mathrm H  \,\vec y =\vec b \,.
\label{eq:normal1_Love1}
\end{equation}

\subsection{Love-Current or Zero-Field Side Condition}
The second possibility to enforce the zero-field condition in an exact manner is to set up a system of equations just for the Love currents without any mapping. 
The assumption that all  currents occurring in~\eqref{eq:LoveCP} are Love currents yields\footnote{The resulting operator in~\eqref{eq:Love2} is just an interior Calderón projector, which is employed to ideally enforce a null field inside of $A$.}
\begin{equation}
\bm{0}
=
\begin{bmatrix} -\frac{1}{2}\bm{\mathcal I} -  \bm{\mathcal K}  & \bm{\mathcal{T}}\\ 
                         -   \bm{\mathcal{T}}                   & -\frac{1}{2}\bm{\mathcal I} -  \bm{\mathcal K}  \end{bmatrix}
\begin{bmatrix} Z_0\bm j_\mathrm{L} \\ \bm m_\mathrm{L} \end{bmatrix}
\,.\label{eq:Love2}
\end{equation}
The equation is utilized as a \ac{SC} 
and no mapping back to the current unkowns is required as for the CP. 
The testing functions can be chosen rather freely. \ac{MFIE}-alike testing with $\bm\beta$-functions leads to
\begin{equation}
\vec{0}
=
\begin{bmatrix} -\frac{1}{2}\mat G_{\bm\beta\bm\beta}  +  \mat K_{\bm\beta}  & \mat T_{\bm\beta}\\ 
                            -\mat T_{\bm\beta} & -\frac{1}{2}\mat G_{\bm\beta\bm\beta}  +  \mat K_{\bm\beta} 
\end{bmatrix}
\begin{bmatrix} Z_0\vec i \\ \vec v \end{bmatrix}
= \mat L_\textsc {sc1} \,\vec x
\label{eq:Love_discr.2}
\end{equation}
with diagonal matrix blocks as known from the classical MFIE. 
Interchanging the magnetic-field with the electric-field equations in~\eqref{eq:Love2} and testing with $\bm\alpha$-function results in 
\begin{equation}
\vec 0 
=
\begin{bmatrix}          
 -\mat T_{\bm\alpha} & -\frac{1}{2}\mat G_{\bm\alpha\bm\beta}  +  \mat K_{\bm\alpha}
 \\
 -\frac{1}{2}\mat G_{\bm\alpha\bm\beta}  +  \mat K_{\bm\alpha}  & \mat T_{\bm\alpha}
\end{bmatrix}
\begin{bmatrix} 
Z_0\vec i
 \\ 
 \vec v \end{bmatrix}
= \mat L_\textsc {sc2} \,\vec x\label{eq:Love_discr.3}
\end{equation}
with the well-tested \ac{EFIE} matrix blocks $\mat T _{\bm \alpha}$, as known from the classical \ac{EFIE}.

It is common knowledge that the low-order \ac{RWG} discretization of the \ac{EFIE} is very accurate, whereas this is not the case for the \ac{MFIE}. 
This may affect the Love-condition discretization and, thus, the achievable level of zero field inside the \ac{AUT} volume. 
Since the numbers of side conditions in~\eqref{eq:Love_discr.2} or~\eqref{eq:Love_discr.3} are actually twice too many, it is also possible to enforce only $N_\mathrm{un}/2$ of the equations in~\eqref{eq:Love_discr.2} or~\eqref{eq:Love_discr.3} with the drawback of interior resonances or, in order to avoid interior resonances, to  combine~\eqref{eq:Love_discr.2} and~\eqref{eq:Love_discr.3} in the manner of a combined-field IE (CFIE) approach. 
The most comprehensive approach with equal weighting of~\eqref{eq:Love_discr.2} and~\eqref{eq:Love_discr.3} and of the electric field and magnetic field leads to
\begin{equation}
\vec 0 
=
\begin{bmatrix}          
 \mat I & \mat I
\end{bmatrix}
\big( \mat L_\textsc {sc1}+\mat L_\textsc {sc2} \big)\,\vec x=\mat L_\textsc {sc3}\,\vec x\label{eq:Love_discr.5}
\end{equation}
with the identity matrices $\mat I\in\mathbb{R}^{N_\mathrm{un}\times N_\mathrm{un}}$.

\section{Scaling of the Zero-Field Side Condition\label{sec:scale}} 
The proper scaling of a \ac{SC} with respect to the forward operator is a challenging task. 
A detailed investigation is necessary to arrive at a problem-independent method. 
In literature, the weighting is never discussed in detail with respect to \ac{NFFFT}s. Either no satisfying solution is proposed in previously reported \ac{NFFFT}s with a SC matrix~\cite{Paulus_comparison_2019}, or the L-curve method is mentioned to find the optimal weighting factor~\cite{Jorg2010,HansenLCurve,Kornprobst2019Love}, which does not seem to be very practical.
The L-curve method means that the (Tikhonov-regularized) inverse problem is solved for a wide range of weighting factors for the side constraint. 
For a weak weighting of the side constraint, this yields a small residual, but a sharp increase of the residual is observed at some value of an increased weighting factor. 
This gives a typically L-shaped curve if the norm of the residual is plotted versus the norm of the regularization term.
The desired solution is found at the knee of the L. 
We perform a similar analysis in the following with the goal of determining an almost problem-independent weighting factor.

In order to find a suitable weighting, we have to discuss the conditioning,\!\footnote{The conditioning of a singular matrix is just related to the nonzero \ac{SV}s of a (truncated) \ac{SVD}.} and hence the \aclp{SVD} (\acsu{SVD}s) of the two matrices under consideration, the forward operator $\mat A$ and the \ac{SC} $\mat L _{\textsc {sc}i}$. 
Both matrices are expected to exhibit a null space. The formulation with electric and magnetic current unknowns is ill-posed per se as mentioned in Section III. Both Love constraint equations in~\eqref{eq:Love_discr.2} and~\eqref{eq:Love_discr.3}
\begin{equation}
\mat L_{\textsc {sc}i}\, \vec x = \vec 0
\end{equation}
must also exhibit a (discretization-limited) nullspace which contains the sought solution. 
All non-Love currents are filtered out  by nonzero \ac{SV}s. 
It is worth noting that a wrong scaling does not only influence the conditioning of the system matrix, but also determines how accurately the inverse problem on the one hand and the \ac{SC} on the other hand are solved for a certain residual stopping threshold. 
It might even occur that the weighting of the \ac{SC} determines the achievable residual threshold.

At this point, we have to distinguish between the \ac{NRE} and the \ac{NEE}. 
The \ac{NRE} preserves the nullspace of $\mat A$ and an additional regularization is both possible and necessary for a unique solution. 
This is a starting point for a meaningful regularization by the Love \ac{SC} to eliminate the null-space of $\mat A$.
The \ac{SC} receives the scaling factor $\xi$, yielding the systems of equations
\begin{equation}
\begin{bmatrix}
\mat A \\
\sqrt\xi\mat L_{\textsc {sc}i}
\end{bmatrix}
\vec x = 
\begin{bmatrix}
\vec b \\
\vec 0
\end{bmatrix}\label{eq:forward_sideconstraint}
\end{equation}
and, subsequently, we attain the modified \ac{NRE}
\begin{equation}
\big [
\mat A^\mathrm{H} \mat A
+
\xi\mat L_{\textsc {sc}i}^\mathrm{H} \mat L_{\textsc {sc}i}^{\vphantom{\mathrm H}}
\big ]
\,
\vec x= 
\mat A^\mathrm{H} 
\vec b \label{eq:scweight}
\end{equation}
augmented by a Tikhonov regularization term.

In the \ac{NEE}, the ill-posedness of the current unknowns $\vec x$ is already mitigated. 
Employing the \ac{SC} as a regularization term is not required anymore and is also not required for a unique solution.
Nevertheless, having~\eqref{eq:forward_sideconstraint} in mind, the formulation of the \ac{NEE} including a \ac{SC} is still possible as discussed in~\cite{Kornprobst2019Love}. Further detailed investigations on the SC weighting  are carried out for the \ac{NRE} only and should be easily transferable to the \ac{NEE}. 

\subsection{Side Condition Scaling for the NRE}
The L-curve approach~\cite{HansenLCurve} shows high computational cost and is highly problem-specific. 
Hence, we propose a weighting based on the \ac{SVD} properties of $\mat A$ and $\mat L_{\textsc {sc}i}
$, which can be stated for any measurement setup.
Employing an iterative solver with residual-limited accuracy, the smaller \ac{SV}s are more likely to be “ignored” dependent on the stopping threshold. 
Thereby, the key is that neither the relevant SVs of the forward operator nor of the Love \ac{SC} are neglected. 
The decay of the \ac{SV}s of $\mat A$ is much stronger than for $\mat L_{\textsc {sc}i}$, since the observability of the various propagating modes decreases with the measurement or observation distance, whereas the observation distance for the Love condition is zero. 

\begin{figure}[t]
\centering
\includegraphics{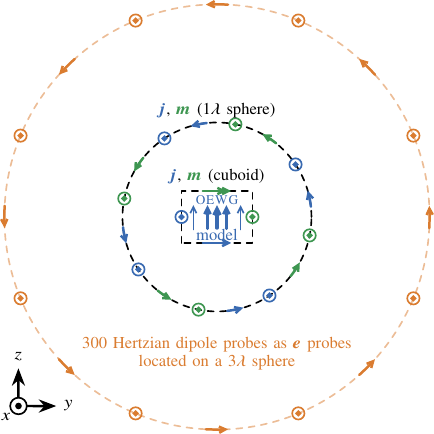}
\caption{Setup of synthetic measurements with Hertzian dipole probes and reconstruction surfaces for equivalent electric and magnetic surface currents. \label{fig:setup1}}
\end{figure}%

\begin{figure}[t]
\centering
\includegraphics{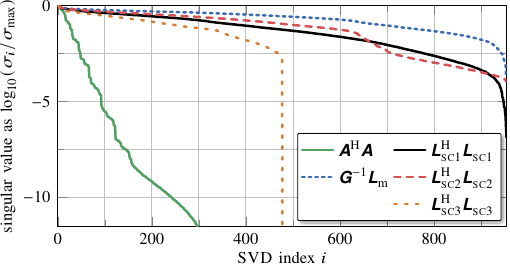}
\caption{Comparison of \ac{SV} spectra of the forward operator and the differently tested Love \ac{SC}s. 
\label{fig:svd1}
}
\end{figure}%

For an empirical study, we employ a small synthetic example, see Fig.~\ref{fig:setup1}. 
A dipole model of an \ac{OEWG}, size $\lambda/4\times\lambda/2\times\lambda/4$, is employed to generate $300$ ideal measurements with Fibonacci sampling~\cite{Fibonacci} on a sphere with radius $3\lambda$. 
The source reconstruction mesh is $\lambda/2\times3\lambda/4\times\lambda/2$ in size and features $477$ \ac{RWG} electric and magnetic current unknowns each.
In Fig.~\ref{fig:svd1}, the \ac{SV} spectra of the forward operator $\mat A$, of the \ac{SC} matrices $\mat L_{\textsc {sc}1}$ (SC1, MFIE-alike), $\mat L_{\textsc {sc}2}$ (SC2, EFIE-alike), $\mat L_{\textsc {sc}3}$ (SC3, CFIE-alike), and  of the \ac{CP} mapping matrix $\mat L_\mathrm m$ are shown. 
While the matrix $\mat A$ exhibits a clear null space after $N_\mathrm{ms}=300$ and a strong decay beforehand, the \ac{SV} decay of the Love-current conditions is weaker. 
Interior and exterior solutions are clearly separated with the SC3-constraint (drop of SVs to zero after $N_\mathrm{un}/2=477$), but this is not clearly observed with the SC1- and SC2-constraints. Here, the discretization errors of the overdetermined sets of equations do not allow one to clearly separate the non-radiating from the strongly evanescent modes. 
The \ac{SV}s are interpreted as follows. 
For the \ac{CP}, Love currents are found at large \ac{SV}s, while for the \ac{SC}s, Love currents are found for the small, close to vanishing,  SVs. 

This small example is mostly intended to be an illustration of the fact that the decay of the SV spectrum of $\mat A$ is steeper than of the side constraint spectra. 
This key property stems from the different interaction distances between the sources and the observation locations in the forward operator and in the side constraint operators, respectively. 
 Larger and more realistic test cases follow in Sections VI and VII.

\begin{figure}[tp]
 \centering%
  \subfloat[]{%
 \includegraphics[]{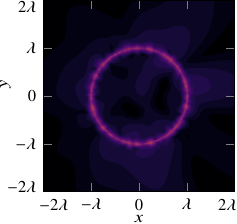}%
   }\hfill%
  \subfloat[]{%
 \includegraphics[]{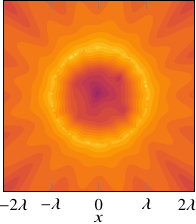}%
   }\hfill%
  \subfloat{%
 \includegraphics[]{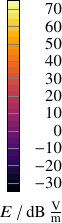}%
   }%
 \caption{Absolute electric field of singular vectors corresponding to 
 (a) $\sigma_\mathrm{max}$ of $\mat L^{\mathrm{H}}_{\textsc {sc}2}\mat L^{\protect\phantom{\mathrm{H}}}_{\textsc {sc}2}$ (non-Love current, evanescent) and to  
 (b) $\sigma_\mathrm{min}$ of $\mat P^{\protect\phantom{\mathrm{H}}}_\matelem{\!A}\mat L^{\mathrm{H}}_{\textsc {sc}2}\mat L^{\protect\phantom{\mathrm{H}}}_{\textsc {sc}2} \mat P^{\protect\phantom{\mathrm{H}}}_\matelem{\!A}$ (Love current, strongly radiating).
\label{fig:fields} }
\end{figure}

Since the \ac{AUT} box with dimensions $3\lambda/4\times\lambda/2\times\lambda/2$ cannot provide a particularly strong field suppression inside, we repeat the same investigation with a sphere with radius $1\lambda$ and about 13\,000 electric and magnetic current unknowns each. 
Only the first $300$ \ac{SV}s and singular vectors are computed~\cite{RohklinSVD}, since the smallest SVs cannot be evaluated realistically due to the matrix size. 
The spectra (not shown) look very similar to Fig.~\ref{fig:svd1} with the only difference of a smaller decay of the \ac{SV}s of $\mat A$, since more modes are excitable on the enlarged equivalent surface\,---\,however, still steeper than the SV decay of the side constraints of course. 
The spherical AUT hull is employed to illustrate the meanings of the SVs in the SC.
The electric field  in a cut plane is evaluated for the singular vector related to the largest SV of the SC2 matrix in Fig.~\ref{fig:fields}(a).
For the smallest SVs, we  project the matrices into the image of $\mat A$ by utilizing the projection matrix $\mat {P}^{\phantom{\mathrm H}}_{\!\matelem A}=\mat A^+\mat A$, i.e., the fields of the singular vector corresponding to the smallest SV of 
$\mat {P}^{\phantom{\mathrm H}}_{\!\matelem A}\mat{L}^\mathrm H_{\textsc {sc}2} \mat L_{\textsc {sc}2}^{\phantom{\mathrm H}}\mat {P}^{\phantom{\mathrm H}}_{\!\matelem A}$
are evaluated in Fig.~\ref{fig:fields}(b).\!\footnote{After the projection into $\mathrm{im}\,\mat A$, there are only $N_\mathrm{ms}\!=\!300$ non-zero SVs instead of $N_\mathrm{un}$ SVs. The projector $\mat {P}^{\phantom{\mathrm H}}_{\!\matelem A}$ suppresses non-observable (non-radiating) components at the measurement locations.}
The radiated fields are rather strong and a certain suppression of the interior field is observed.
Overall, a separation of strongly and weakly radiating currents is observed for the largest and smallest \ac{SV}s, respectively. 
For the CP, the meanings of maximum and minimum SVs are interchanged.

Note that all considered singular vectors are orthonormal, i.e., their possible influence on the norm $\lVert\vec x\rVert_2$ is the same and non-Love currents are suppressed by the Love condition \emph{only at a suitable weighting}. A weak weighting will have no effect on the inverse problem solution, and a strong weighting limits the achievable solver residual threshold, which has a similar effect as an earlier solver termination.

The question remains as how to incorporate these insights in a meaningful way into the transformation process appropriate for any kind of measurement.
From the \ac{SVD}s, it is clear that an equal weighting according to the maximum \ac{SV}s\,---\,as shown in Fig.~\ref{fig:svd1}\,---\,will lead to a dominant \ac{SC} and a poor solution quality of the inverse problem. 
With the knowledge of the achievable accuracy of the algorithm and the measurement setup,\!\footnote{The \ac{FIAFTA} itself is limited by  numerical errors controllable by the \ac{MLFMM} accuracy settings; 
measurements have inherent limitations by the \ac{SNR} and other measurement errors.} one can choose a weighting of the \ac{SC} at a somewhat larger ratio than the expected accuracy. 
The accuracy, e.g., as a measure for \ac{SNR}, can be quantified at the maximum of the measured \ac{AUT} \ac{NF} as
\begin{equation}
\mathit{SNR} = \frac{\lVert\bm\epsilon_\textsc{oe}\rVert_2/\sqrt{N_\mathrm{ms}}}{\max\limits _{k\in [0~N_\mathrm{ms}]} \lvert[\vec b]_k\rvert}
\end{equation}
scaled according to the number of observation points $N_\mathrm{ms}$ and assuming $\lVert\bm\epsilon_\textsc{oe}\rVert_2$ is known.
The relative $\ell_2$-norm of the \ac{OE} in the \ac{NF} is estimated as the \ac{RD} $\epsilon_\textsc{rd}$.
As discussed in Section II, the \ac{RD} of the retrieved solution $\epsilon_\textsc{rd}$ is commonly a bit smaller than $\epsilon_\textsc{oe}$ since parts of the \ac{OE} are inevitably attributed to the solution of the inverse problem. 
Measurement errors (echoes, positioning uncertainties, etc.) may increase $\epsilon_\textsc{oe}$.
These quantities, and the eventually achievable $\epsilon_\textsc{oe}$, are commonly known for a measurement setup and determine the iterative solver stopping threshold or the \ac{SVD} truncation criterion for direct solvers.

The correct weighting of the \ac{SC} is now apparent: The Love current \ac{SV}s\,---\,i.e., the smallest \ac{SV}s of the \ac{SC}\,---\,have to disappear below $\epsilon_\textsc{oe}^2$, and the non-Love current SVs have to be located above $\epsilon_\textsc{oe}^2$. 
With a suitable iterative solver termination criterion based on $\epsilon_\textsc{oe}$, the non-Love parts in the solution are effectively suppressed.
Certainly, the full \ac{SVD} is computationally too costly, but a normalization relative to the largest \ac{SV} is feasible and also reasonable, if we keep in mind that the decay of the SVs related to the Love-SC is considerably slower than the decay of the SVs related to $\mat A$.
Estimating the largest \ac{SV} is easily done with a few so-called power iterations. 
A repeated evaluation of the matrix-vector product gives a good estimate of the largest \ac{SV}
\begin{equation}
\hat\sigma_{\matelem B,\mathrm{max}} = \frac{\vec x ^\mathrm H \mat B ^k \vec x}{\vec x ^\mathrm H \mat B ^{k-1} \vec x }
 \,,
\end{equation}
assuming a square and Hermitian matrix $\mat B$.
This method converges rather fast, for a reasonable residual below $10^{-1}$ typically with $k<5$. 
Other possibilities include Arnoldi iterations or a Rayleigh quotient with the starting vector $ \mat A^\mathrm H\vec b$.

\begin{figure}[tp]
\centering
\includegraphics{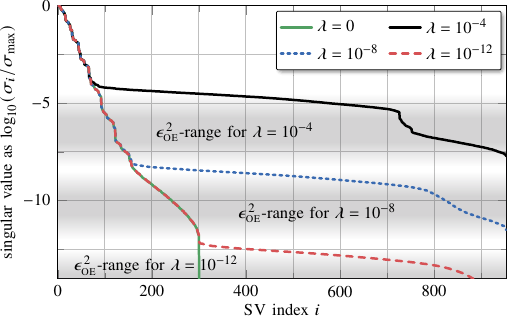}
\caption{Analysis of the  weighting $\lambda$ by a look at the \ac{SV}s of~\eqref{eq:scweight}. Gray areas approximately show possible $\epsilon_\textsc{oe}$ ranges for the given weights.\label{fig:svd2}}
\end{figure}

In order to attain a normalization according to the largest \ac{SV}s of $\mat A$ and the Love-SC, respectively, (as seen in Fig.~\ref{fig:svd1}) we employ the (estimated) ratio 
\begin{equation}
\zeta=
\frac
{\hat{\sigma}_{\matelem A^\mathrm H\matelem A}}{\hat{\sigma}_{\matelem L \mathrm {sc}i^\mathrm H\matelem L \mathrm {sc}i}}\,.
\end{equation}
The normalization has to be done in a way that, first, the larger SVs of $\mat A$ influence the reconstruction and, second, the iterative solver termination threshold at $\epsilon_\textsc{oe}$ coincides with the desired SV threshold within the SV spectrum of the Love-SC. This is achieved by choosing the scaling factor $\xi$ according to~\eqref{eq:scweight} as
\begin{equation}
\xi=
\zeta\lambda\,,\quad\mathrm{with}~1>\lambda>\epsilon_\textsc{oe}^2\,.\label{eq:xi}
\end{equation}
For the studied example of an \ac{OEWG}, the \ac{SV}s of the combined operator according to~\eqref{eq:scweight} of several values of $\lambda$ and for an \ac{EFIE}-alike \ac{SC} are shown in Fig.~\ref{fig:svd2}.
An $\epsilon_\textsc{oe}^2$ range is depicted for each $\lambda$ value. The corresponding $\epsilon_\textsc{oe}$ value is determined as follows. 
Based on the \ac{NRE} including an error vector $\bm\epsilon_\textsc{oe}$ within $\vec b$, we recognize that $\mat A ^\mathrm H$ is multiplied to both the forward operator and the error vector. 
Hence, the $\epsilon_\textsc{oe}$ range in Fig.~\ref{fig:svd2} has to be considered. 
We assume that, for the discussed scenario,
\begin{equation}
\lambda = \big(10^1\ldots10^3\big)\epsilon_\textsc{oe}^2 
\end{equation}
is a good choice. For larger scenarios, the upper bound $\lambda$ is shifted to slightly larger values. 
In the results section, it is demonstrated that this scaling is indeed meaningful. 
Analyzing this relation in depth would require  the L-curve method.

\section{Results with Synthetic Near-Field Data} %
\subsection{Transformation Results for the Small Synthetic OEWG}
For the discussed \ac{OEWG} example, we consider the equivalent source types listed in Tab.~\ref{tab:acs}(b),
the various types of surface current densities, but also\,---\,for completeness regarding (surface) source representations\,---\,\ac{SH} and \ac{D-SH} sources. For the latter, the MLFMM-octree is built for the AUT hull and SH expansions are employed in the non-empty lowest-level boxes with a size of $0.15$ wavelengths.

\begin{figure}[tp]
 \centering%
  \subfloat[]{%
 \includegraphics[]{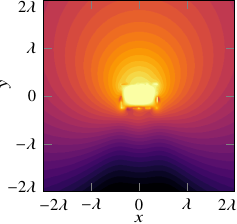}%
   }\hfill%
  \subfloat[]{%
 \includegraphics[]{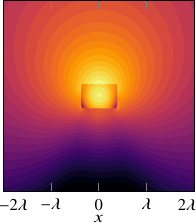}%
   }\hfill%
  \subfloat{%
 \includegraphics[]{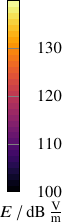}%
   }%
   \par\setcounter{subfigure}{2}
  \subfloat[]{%
 \includegraphics[]{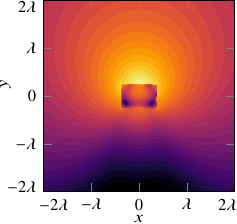}%
   }\hfill%
  \subfloat[]{%
 \includegraphics[]{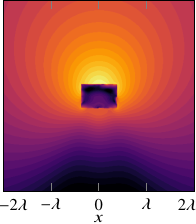}%
   }\hfill%
  \subfloat{%
 \includegraphics[]{fig6x}%
   }%
   \par\setcounter{subfigure}{4}
  \subfloat[]{%
 \includegraphics[]{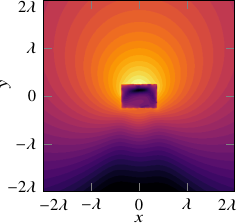}%
   }\hfill%
  \subfloat[]{%
 \includegraphics[]{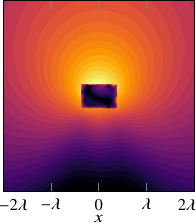}%
   }\hfill%
  \subfloat{%
 \includegraphics[]{fig6x}%
   }%
 \caption{Reconstructed electric field in a cut plane of the (a) J -- NEE, (b) JM -- NEE, (c) CS -- NEE, and (d) CP -- NEE solutions with $\epsilon_\mathrm{res}=10^{-2}$ termination criterion as well as (e) SC1 -- NRE ($\lambda=10^2$) and (f) SC2 -- NRE  ($\lambda=10^1$) solutions with $\epsilon_\mathrm{res}=10^{-4}$ termination criterion.
\label{fig:fields2} }
\end{figure}

White Gaussian noise with $\epsilon_\textsc{oe}=10^{-2}$  is added to the ideal synthetic data. 
All the solvers except the SC ones have been analyzed for the small cube mesh with regard to their properties and the results are summarized in Tab.~\ref{tab:side-weight}.
\colorlet{vgood}{green!45!white}%
\colorlet{good}{blue!35!white}%
\colorlet{bad}{orange!95!black}%
\colorlet{vbad}{red!88!black}%
\definecolor{neut}{gray}{.78}%
\def\vg{\cellcolor{vgood}}%
\def\gd{\cellcolor{good}}%
\def\bd{\cellcolor{bad}\leavevmode\color{white}}%
\def\vb{\cellcolor{vbad}\leavevmode\color{white}}%
\def\nt{\cellcolor{neut}}%
\begin{table}[tp]
\caption{Comparison of various source types for the two NEs, $\epsilon_\textsc{oe}=10^{-2}$. (\textup{a}) NRE with  $\epsilon_\mathrm{res}=10^{-4}$. (\textup{b})~NRE with $\Deltaup\epsilon_\mathrm{res}=0.99$. \mbox{(\textup{c})~NEE with $\epsilon_\mathrm{res}=10^{-2}$. (\textup{d})~NEE with $\Deltaup\epsilon_\mathrm{res}=0.99$.\label{tab:side-weight}}}
\centering
\subfloat[]{
\begin{tabular}{L{17.5mm} R{4mm}R{10.5mm}R{9mm}R{10.5mm}R{10.5mm}}
\toprule[1pt]
source type&   $N_\mathrm{it}$ &  $\epsilon_\mathrm{res}$ &$\epsilon_\textsc{rd}$/$\epsilon_\textsc{oe}$ & $\epsilon_\mathrm{\textsc{zf},avg}$ & $\epsilon_\mathrm{\textsc{ff},max}$ \\
\cmidrule[0.5pt](lr){1-1} \cmidrule[0.5pt](lr){2-2}  \cmidrule[0.5pt](lr){3-3}  \cmidrule[0.5pt](lr){4-4} \cmidrule[0.5pt](lr){5-5} \cmidrule[0.5pt](lr){6-6}
SH -- NRE        &\vg  $  5$  &\nt  $6.1\cdot10^{-5}$  &\bd  $0.64$  &\nt---\phantom{\,dB}&\bd  $-43.2$\,dB  \\[0.2ex]
D-SH             &\gd  $ 24$  &\nt  $9.9\cdot10^{-5}$  &\bd  $0.77$  &\nt---\phantom{\,dB}&\bd  $-44.6$\,dB  \\[0.2ex]
J -- NRE         &\bd  $ 41$  &\nt  $7.8\cdot10^{-5}$  &\vg  $0.92$  &\bd    $8.5$\,dB    &\vg  $-48.4$\,dB  \\[0.2ex]
M -- NRE         &\gd  $ 32$  &\nt  $9.8\cdot10^{-5}$  &\gd  $1.02$  &\bd    $2.0$\,dB    &\bd  $-44.9$\,dB  \\[0.2ex]
JM -- NRE        &\gd  $ 32$  &\nt  $8.2\cdot10^{-5}$  &\vg  $0.94$  &\bd    $0.0$\,dB    &\vg  $-48.6$\,dB  \\[0.2ex]
CS -- NRE        &\gd  $ 32$  &\nt  $9.2\cdot10^{-5}$  &\vg  $0.94$  &\gd  $ -5.8$\,dB    &\vg  $-48.2$\,dB  \\[0.2ex]
CP -- NRE        &\gd  $ 31$  &\nt  $9.6\cdot10^{-5}$  &\vg  $0.99$  &\gd  $-16.6$\,dB    &\vg  $-49.4$\,dB  \\[0.2ex]
WF-CP -- NRE     &\gd  $ 32$  &\nt  $9.9\cdot10^{-5}$  &\vg  $0.98$  &\vg  $-21.1$\,dB    &\vg  $-48.8$\,dB  \\
\bottomrule[1pt]
\end{tabular}
}\\
\subfloat[]{
\begin{tabular}{L{17.5mm} R{4mm}R{10.5mm}R{9mm}R{10.5mm}R{10.5mm}}
\toprule[1pt]
SH -- NRE        &\gd  $ 16$  &\nt $5.0\cdot10^{-8}$  &\bd  $0.64$  &\nt---\phantom{\,dB}&\bd  $-43.2$\,dB  \\[0.2ex]
D-SH             &\vb  $208$  &\nt $1.6\cdot10^{-7}$  &\vb  $0.12$  &\nt---\phantom{\,dB}&\vb  $-40.6$\,dB  \\[0.2ex] 
J -- NRE         &\vb  $134$  &\nt $8.9\cdot10^{-5}$  &\bd  $0.62$  &\vb   $30.4$\,dB    &\bd  $-43.6$\,dB\\[0.2ex]
M -- NRE         &\vb  $139$  &\nt $4.4\cdot10^{-8}$  &\bd  $0.58$  &\vb   $24.1$\,dB    &\bd  $-42.9$\,dB  \\[0.2ex]
JM -- NRE        &\vb  $152$  &\nt $3.8\cdot10^{-8}$  &\vb  $0.45$  &\vb   $40.0$\,dB    &\bd  $-43.4$\,dB  \\[0.2ex]
CS -- NRE        &\vb  $121$  &\nt $1.9\cdot10^{-7}$  &\vb  $0.69$  &\vb  $ 18.1$\,dB    &\bd  $-43.6$\,dB  \\[0.2ex]
CP -- NRE        &\bd  $ 48$  &\nt $3.2\cdot10^{-5}$  &\gd  $0.87$  &\gd  $-11.8$\,dB    &\vg  $-48.0$\,dB  \\[0.2ex]
WF-CP -- NRE     &\bd  $ 48$  &\nt $4.9\cdot10^{-5}$  &\gd  $0.87$  &\gd  $-11.8$\,dB    &\gd  $-46.7$\,dB  \\
\bottomrule[1pt]
\end{tabular}
}\\
\subfloat[]{
\begin{tabular}{L{17.5mm} R{4mm}R{10.5mm}R{9mm}R{10.5mm}R{10.5mm}}
\toprule[1pt]
SH -- NEE        &\vg  $  2$  &\bd  $6.9\cdot10^{-3}$  &\bd  $0.69$  &\nt---\phantom{\,dB}&\bd  $-43.3$\,dB  \\[0.2ex]
D-SH             &\vg  $  5$  &\vg  $9.7\cdot10^{-3}$  &\vg  $0.97$  &\nt---\phantom{\,dB}&\vb  $-40.4$\,dB  \\[0.2ex]
J -- NEE         &\bd  $ 37$  &\vg  $9.9\cdot10^{-5}$  &\vg  $0.99$  &\bd    $8.5$\,dB    &\gd  $-45.1$\,dB\\[0.2ex]
M -- NEE         &\gd  $ 32$  &\vg  $9.6\cdot10^{-3}$  &\vg  $0.96$  &\bd    $2.0$\,dB    &\gd  $-45.0$\,dB  \\[0.2ex]
JM -- NEE        &\gd  $ 26$  &\vg  $9.7\cdot10^{-3}$  &\vg  $0.97$  &\bd    $0.0$\,dB    &\vg  $-48.0$\,dB  \\[0.2ex]
CS -- NEE        &\gd  $ 28$  &\vg  $9.8\cdot10^{-3}$  &\vg  $0.98$  &\gd  $ -7.7$\,dB    &\vg  $-47.2$\,dB  \\[0.2ex]
CP -- NEE        &\gd  $ 30$  &\vg  $9.9\cdot10^{-3}$  &\vg  $0.99$  &\vg  $-21.2$\,dB    &\vg  $-47.5$\,dB  \\[0.2ex]
WF-CP -- NEE     &\gd  $ 30$  &\vg  $9.9\cdot10^{-3}$  &\vg  $0.99$  &\vg  $-21.6$\,dB    &\vg  $-47.7$\,dB  \\
\bottomrule[1pt]
\end{tabular}
}\\
\subfloat[]{
\begin{tabular}{L{17.5mm} R{4mm}R{10.5mm}R{9mm}R{10.5mm}R{10.5mm}}
\toprule[1pt]
SH -- NEE        &\vg  $  7$  &\bd  $6.4\cdot10^{-3}$  &\bd  $0.64$  &\nt---\phantom{\,dB}&\bd  $-43.3$\,dB  \\[0.2ex]
D-SH             &\gd  $ 14$  &\bd  $8.0\cdot10^{-3}$  &\bd  $0.80$  &\nt---\phantom{\,dB}&\gd  $-45.3$\,dB  \\[0.2ex]
J -- NEE         &\bd  $ 45$  &\gd  $8.7\cdot10^{-5}$  &\gd  $0.87$  &\bd    $8.6$\,dB    &\vg  $-47.6$\,dB\\[0.2ex]
M -- NEE         &\bd  $ 38$  &\gd  $8.9\cdot10^{-3}$  &\gd  $0.89$  &\bd    $2.1$\,dB    &\vg  $-48.3$\,dB  \\[0.2ex]
JM -- NEE        &\gd  $ 33$  &\vg  $9.2\cdot10^{-3}$  &\vg  $0.92$  &\bd    $5.2$\,dB    &\vg  $-47.6$\,dB  \\[0.2ex]
CS -- NEE        &\gd  $ 34$  &\vg  $9.1\cdot10^{-3}$  &\vg  $0.91$  &\gd  $ -3.7$\,dB    &\vg  $-48.2$\,dB  \\[0.2ex]
CP -- NEE        &\gd  $ 35$  &\vg  $9.5\cdot10^{-3}$  &\vg  $0.95$  &\vg  $-21.3$\,dB    &\vg  $-47.9$\,dB  \\[0.2ex]
WF-CP -- NEE     &\gd  $ 35$  &\vg  $9.5\cdot10^{-3}$  &\vg  $0.95$  &\vg  $-21.7$\,dB    &\vg  $-47.7$\,dB  \\
\bottomrule[1pt]
\end{tabular}
}
\end{table}%
Excellent results are highlighted in 
\sethlcolor{vgood}\hl{bright green}, 
good ones in 
\sethlcolor{good}\hl{blue}, 
worse ones in 
\sethlcolor{bad}{\color{white}\hl{darkish orange}},
and the underwhelmingly poor ones in 
\mbox{\sethlcolor{vbad}{\color{white}\hl{dark red}}.}
Cells without a judgment are 
\sethlcolor{neut}\hl{gray}.
To put these ratings into perspective, the following should be considered. 
An iteration count $N_\mathrm{it}$ below ten is rather good, while hundreds of iterations are too much. 
The solver residual $\epsilon_\mathrm{res}$ only conveys information for the NEE case. 
The normalized \ac{RD} $\epsilon_\textsc{rd}/\epsilon_\textsc{oe}$ has an ideal value of $1$, with lower values indicating over-fitting and larger values indicating a wrong solution. 
The zero field  quality inside $A$ $\epsilon_\mathrm{\textsc{zf},avg}$ is judged by averaging the field inside $A$ at 100 observation points, normalized to the JM-NEE case. 
This is visualized in Fig.~\ref{fig:fields2} for various different solutions.
The relative \ac{FF} error between the reconstructed FF $\bm e$ and the reference~$\bm e_\mathrm{ref}$ reads
\begin{equation}
\epsilon_\mathrm{\textsc{ff},max}=\max\limits_{\vartheta,\varphi}\Big\lVert\frac{\bm e}{\max e_{\vartheta,\varphi}}-\frac{ \bm e_\mathrm{ref}}{\max e_{\vartheta,\varphi,\mathrm{ref}}}\Big\rVert_2\,.
\end{equation}

Iterative solver settings are chosen as $\epsilon_\mathrm{res}=10^{-4}$ for the \ac{NRE} and $\epsilon_\mathrm{res}=\epsilon_\textsc{rd}=10^{-2}$ for the \ac{NEE}. 
We further investigate a somewhat relaxed, relative termination criterion\footnote{The advantage of a relative stopping criterion is that no pre-knowledge about the measurement setup (i.e., knowledge of the OE) is required.} dependent on the iterative solver convergence ratio: the solver stops if the relative residual improvement $\Deltaup\epsilon$ becomes worse than $0.99$ three times in a row. This is demonstrated to work almost as well as the absolute stopping for the \ac{NEE} in Tab.~\ref{tab:side-weight}(d), whereas the \ac{NRE} struggles to prevent overfitting without an absolute stopping, compare Tab.~\ref{tab:side-weight}(a) and (b). 
Overfitting leads to a solution which is numerically contaminated by the null space of $\mat A$. Furthermore, we note that the \ac{NEE} version always converges in fewer iterations than the \ac{NRE} for any specific source type or stopping criterion.

Various values of $\lambda$ for both the SC1 and the SC2 discretizations are investigated in Tab.~\ref{tab:side-weight2}, 
\begin{table}[tp]
\caption{Analysis of Love \ac{SC} weighting $\lambda$, NRE, $\epsilon_\textsc{oe}=10^{-2}$. \mbox{(\textup{a})~SC1 (MFIE)  $\epsilon_\mathrm{res}=10^{-4}$.  (\textup{b})~SC1 (MFIE) $\Deltaup\epsilon_\mathrm{res}=0.99$.} \mbox{(\textup{c})~SC2 (EFIE) $\epsilon_\mathrm{res}=10^{-4}$. (\textup{d})~SC2 (EFIE) $\Deltaup\epsilon_\mathrm{res}=0.99$.\label{tab:side-weight2}}}
\centering
\subfloat[]{
\begin{tabular}{L{17.5mm} R{4mm}R{10.5mm}R{9mm}R{10.5mm}R{10.5mm}}
\toprule[1pt]
SC weight&   $N_\mathrm{it}$ &  $\epsilon_\mathrm{res}$ &$\epsilon_\textsc{rd}$/$\epsilon_\textsc{oe}$ & $\epsilon_\mathrm{\textsc{zf},avg}$ & $\epsilon_\mathrm{\textsc{ff},max}$ \\
\cmidrule[0.5pt](lr){1-1} \cmidrule[0.5pt](lr){2-2}  \cmidrule[0.5pt](lr){3-3}  \cmidrule[0.5pt](lr){4-4} \cmidrule[0.5pt](lr){5-5} \cmidrule[0.5pt](lr){6-6}
$\lambda=10^0\epsilon_\textsc{oe}^2$    &\gd $ 32$  &\nt  $8.8\cdot10^{-5}$  &\vg  $0.94$  &\bd  $-0.23$\,dB  &\vg  $-48.7$\,dB  \\[0.2ex]
$\lambda=10^{0.5}\epsilon_\textsc{oe}^2$&\bd $ 36$  &\nt  $9.9\cdot10^{-5}$  &\vg  $0.92$  &\bd   $-1.0$\,dB  &\vg  $-49.2$\,dB  \\[0.2ex]
$\lambda=10^1\epsilon_\textsc{oe}^2$    &\bd $ 40$  &\nt  $9.3\cdot10^{-5}$  &\gd  $0.89$  &\gd  $ -9.7$\,dB  &\vg  $-49.0$\,dB  \\[0.2ex]
$\lambda=10^{1.5}\epsilon_\textsc{oe}^2$&\bd $ 39$  &\nt  $9.6\cdot10^{-5}$  &\vg  $0.90$  &\vg  $-16.7$\,dB  &\vg  $-49.4$\,dB  \\[0.2ex]
$\lambda=10^2\epsilon_\textsc{oe}^2$    &\bd $ 39$  &\nt  $9.4\cdot10^{-5}$  &\vg  $0.92$  &\vg  $-18.3$\,dB  &\vg  $-48.1$\,dB  \\[0.2ex]
$\lambda=10^3\epsilon_\textsc{oe}^2$    &\bd $ 49$  &\nt  $9.9\cdot10^{-5}$  &\vg  $0.98$  &\vg  $-18.2$\,dB  &\vb  $-41.4$\,dB  \\[0.2ex]
$\lambda=10^4\epsilon_\textsc{oe}^2$    &\vb $101$  &\nt  $9.7\cdot10^{-5}$  &\vb  $2.33$  &\vg  $-16.9$\,dB  &\vb  $-41.4$\,dB  \\
\bottomrule[1pt]
\end{tabular}
}\\
\subfloat[]{
\begin{tabular}{L{17.5mm} R{4mm}R{10.5mm}R{9mm}R{10.5mm}R{10.5mm}} 
\toprule[1pt]
$\lambda=10^0\epsilon_\textsc{oe}^2$    &\vb $225$  &\nt  $1.1\cdot10^{-9}$  &\bd  $0.76$  &\gd  $-11.9$\,dB  &\gd  $-44.7$\,dB  \\[0.2ex]
$\lambda=10^{0.5}\epsilon_\textsc{oe}^2$&\vb $226$  &\nt  $1.4\cdot10^{-9}$  &\bd  $0.78$  &\gd  $-14.5$\,dB  &\gd  $-45.3$\,dB  \\[0.2ex]
$\lambda=10^1\epsilon_\textsc{oe}^2$    &\vb $238$  &\nt  $1.2\cdot10^{-9}$  &\bd  $0.80$  &\vg  $-16.2$\,dB  &\gd  $-45.9$\,dB  \\[0.2ex]
$\lambda=10^{1.5}\epsilon_\textsc{oe}^2$&\vb $249$  &\nt  $1.2\cdot10^{-9}$  &\gd  $0.82$  &\vg  $-17.2$\,dB  &\vg  $-46.3$\,dB  \\[0.2ex]
$\lambda=10^2\epsilon_\textsc{oe}^2$    &\vb $268$  &\nt  $1.2\cdot10^{-9}$  &\gd  $0.84$  &\vg  $-17.5$\,dB  &\gd  $-45.9$\,dB  \\[0.2ex]
$\lambda=10^3\epsilon_\textsc{oe}^2$    &\vb $326$  &\nt  $1.7\cdot10^{-9}$  &\gd  $1.03$  &\vg  $-17.2$\,dB  &\bd  $-43.2$\,dB  \\[0.2ex]
$\lambda=10^4\epsilon_\textsc{oe}^2$    &\vb $409$  &\nt  $3.3\cdot10^{-9}$  &\vb  $2.35$  &\vg  $-16.5$\,dB  &\vb  $-38.2$\,dB  \\
\bottomrule[1pt]
\end{tabular}
}\\
\subfloat[]{
\begin{tabular}{L{17.5mm} R{4mm}R{10.5mm}R{9mm}R{10.5mm}R{10.5mm}}
\toprule[1pt]
$\lambda=10^{-1}\epsilon_\textsc{oe}^2$ &\gd $ 32$  &\nt  $8.3\cdot10^{-5}$  &\vg  $0.94$  &\bd  $-0.03$\,dB  &\vg  $-48.6$\,dB  \\[0.2ex]
$\lambda=10^0\epsilon_\textsc{oe}^2$    &\gd $ 32$  &\nt  $9.2\cdot10^{-5}$  &\vg  $0.93$  &\bd  $-0.33$\,dB  &\vg  $-48.7$\,dB  \\[0.2ex]
$\lambda=10^{0.5}\epsilon_\textsc{oe}^2$&\bd $ 40$  &\nt  $9.9\cdot10^{-5}$  &\gd  $0.89$  &\bd   $-3.7$\,dB  &\vg  $-49.2$\,dB  \\[0.2ex]
$\lambda=10^1\epsilon_\textsc{oe}^2$    &\bd $ 40$  &\nt  $9.5\cdot10^{-5}$  &\gd  $0.89$  &\vg  $-21.5$\,dB  &\vg  $-49.0$\,dB  \\[0.2ex]
$\lambda=10^{1.5}\epsilon_\textsc{oe}^2$&\bd $ 39$  &\nt  $8.1\cdot10^{-5}$  &\vg  $0.91$  &\vg  $-15.7$\,dB  &\vg  $-49.4$\,dB  \\[0.2ex]
$\lambda=10^2\epsilon_\textsc{oe}^2$    &\bd $ 38$  &\nt  $9.8\cdot10^{-5}$  &\vg  $0.94$  &\gd  $-14.8$\,dB  &\vg  $-48.1$\,dB  \\[0.2ex]
$\lambda=10^3\epsilon_\textsc{oe}^2$    &\bd $ 68$  &\nt  $9.9\cdot10^{-5}$  &\bd  $1.18$  &\gd  $-12.5$\,dB  &\vb  $-41.4$\,dB  \\[.2ex]
$\lambda=10^4\epsilon_\textsc{oe}^2$    &\vb $128$  &\nt  $9.7\cdot10^{-5}$  &\vb  $3.63$  &\gd  $-11.6$\,dB  &\vb  $-27.0$\,dB  \\
\bottomrule[1pt]
\end{tabular}
}\\
\subfloat[]{
\begin{tabular}{L{17.5mm} R{4mm}R{10.5mm}R{9mm}R{10.5mm}R{10.5mm}}
\toprule[1pt]
$\lambda=10^{-1}\epsilon_\textsc{oe}^2$ &\vb $236$  &\nt  $1.5\cdot10^{-9}$  &\bd  $0.72$  &\bd  $-2.1$\,dB   &\gd  $-44.0$\,dB  \\[0.2ex]
$\lambda=10^0\epsilon_\textsc{oe}^2$    &\vb $262$  &\nt  $1.4\cdot10^{-9}$  &\bd  $0.76$  &\gd  $-9.3$\,dB   &\gd  $-44.9$\,dB  \\[0.2ex]
$\lambda=10^{0.5}\epsilon_\textsc{oe}^2$&\vb $266$  &\nt  $1.5\cdot10^{-9}$  &\bd  $0.78$  &\gd  $-10.9$\,dB  &\gd  $-45.7$\,dB  \\[0.2ex]
$\lambda=10^{1}\epsilon_\textsc{oe}^2$  &\vb $304$  &\nt  $1.5\cdot10^{-9}$  &\gd  $0.81$  &\gd  $-11.6$\,dB  &\vg  $-46.6$\,dB  \\[0.2ex]
$\lambda=10^{1.5}\epsilon_\textsc{oe}^2$&\vb $309$  &\nt  $2.2\cdot10^{-9}$  &\gd  $0.84$  &\gd  $-11.8$\,dB  &\vg  $-46.6$\,dB  \\[0.2ex]
$\lambda=10^2\epsilon_\textsc{oe}^2$    &\vb $329$  &\nt  $1.6\cdot10^{-9}$  &\gd  $0.88$  &\gd  $-11.7$\,dB  &\vg  $-46.8$\,dB  \\[0.2ex]
$\lambda=10^3\epsilon_\textsc{oe}^2$    &\vb $360$  &\nt  $4.1\cdot10^{-9}$  &\bd  $1.19$  &\gd  $-11.6$\,dB  &\bd  $-42.0$\,dB  \\[0.2ex]
$\lambda=10^4\epsilon_\textsc{oe}^2$    &\vb $392$  &\nt  $1.0\cdot10^{-8}$  &\vb  $3.60$  &\gd  $-11.5$\,dB  &\vb  $-27.2$\,dB  \\
\bottomrule[1pt]
\end{tabular}
} 
\end{table}%
again for the same absolute and relative stopping criteria.
We observe that the \ac{SC} with increasing weighting 
i) slows down the iterative solver (undesired), 
ii) increases the \ac{FF} error if the weighting becomes too strong (undesired), and
iii) increases the \ac{RD} (up to a certain limit, desirable).
The RD limitation is observed even though the solver performs hundreds of iterations and reaches a residual of $10 ^{-9}$\,---\,if the correct weight of about $\lambda=10 ^1\ldots10^2$ is set in Tab.~\ref{tab:side-weight2}(b) and (d).
The same is observed, to a certain degree, for the CP and for the WF-CP  in Tab.~\ref{tab:side-weight}(b). 

Overall, the only ensured and meaningful effect of enforcing a Love-current solution is a zero field inside $A$, see Figs.~\ref{fig:fields2}(e) and (f). 
The second effect of influencing the iterative solver residual threshold is achieved much more conveniently\,---\,and at a much lower computation cost\,---\,with an ambigous JM solution or a unique \ac{CS} solution by choosing an appropriate stopping criterion for the selected normal equation, with a clear preference for the NEE.

Interestingly, no significant differences in the solution quality between MFIE- and EFIE-alike discretizations of the Love SC can be identified\,---\,\emph{while this classical MFIE discretization causes accuracy problems for scattering problems}. Even the zero-field quality is comparable, with slight advantages for the MFIE-alike SC1. A similar lack of influence is observed for the CP, where the potentially accuracy-improved WF-CP does not perform better in any of the considered measures.

The \ac{SH} approach shows the fastest iterative solver convergence, however, also an about 3\,dB to 8\,dB larger FF error. However, this is by no means a \emph{fair} comparison\footnote{Since the measurement distance of this example is quite large (beneficial for SH), the iterative solver convergence of the SH NFFFT is extremly fast. However, the SH solution has the drawback that it cannot offer the same diagnostic information and source localization as an equivalent current approach and, thus, exhibits larger NF and FF errors.}\!\!. It is listed for the sake of completeness.
The second-fastest source type is \ac{D-SH}, again without the full diagnostic capabilities.
The fastest surface-current solutions are obtained either with ambiguous JM or unique CS currents. The use of the NEE speeds up the solution process from 32 to 26 and 28 iterations, respectively.
Furthermore, we observe that the JM and CS solutions are among the most accurate ones.

In this investigation, the JM-NEE and CS-NEE are the fastest-convergent (i.e., best conditioned) and most accurate (in the NF and FF) surface-current formulations. 
Hence, JM or  CS currents with the NEE seem to be the most reasonable choice for best accuracy and best conditioning.
Diagnostics capabilities can be easily enhanced in the  post-processing~\cite{Kornprobst2019Love}.

\begin{figure}[tp]
\centering
\subfloat[]{\includegraphics[]{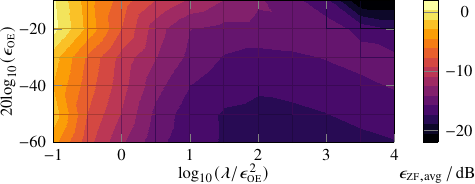}}\\
\subfloat[]{\includegraphics[]{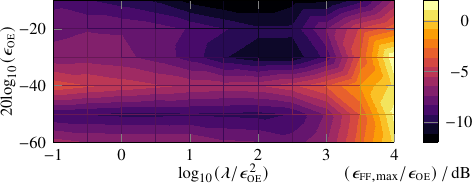}}
\caption{Love SC1 weighting analysis regarding (a) zero-field quality and (b) maximum FF error for varying OE levels, OEWG simulation model.
\label{fig:LoveSCweight}}
\end{figure}

Finally, we analyze the weighting of the Love SC1 for different SNRs in Fig.~\ref{fig:LoveSCweight}.
An acceptably suppressed interior field is observed with weightings larger than $\lambda=10 ^{1}\epsilon_\textsc{oe}^2$, and the reconstructed FF deteriorates with a too strong weighting above $\lambda=10 ^{3}\epsilon_\textsc{oe}^2$. 
Note that the FF error is given with reference to the OE; the processing gain is therefore included. 
As seen in Tab.~\ref{tab:side-weight2}, a similar behavior is observed for the SC2 variant. 
However, the zero-field quality is more sensitive especially to the choice of the stopping criterion.

\subsection{A More Realistic Antenna Model}

In the following, a simulation model of a DRH400 antenna at 6\,GHz is considered~\cite{DRH400}. 
The integral equation solver of the simulation software Feko was employed to generate synthetic \ac{NF} data ($3754$ measurement samples) for a spiral scan with $20\si{\percent}$ oversampling in relation to the minimum sphere of the \ac{AUT}~\cite{Feko,Hansen, Bucci}. 
The advantage compared to measurements is that the real solution (i.e., a reference) is known from simulation. 

\begin{figure}[tp]
\centering
\subfloat[]{\includegraphics[]{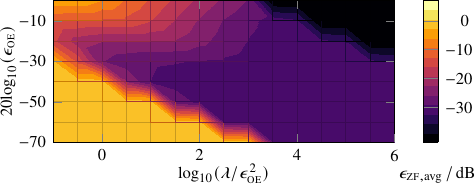}}\\
\subfloat[]{\includegraphics[]{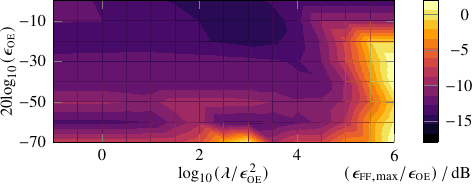}}
\caption{Love SC1 weighting analysis regarding (a) zero-field quality and (b) maximum FF error for varying OE levels, DRH400 simulation model.
\label{fig:LoveSCweight_drh}}
\end{figure}

First, we perform an analysis of the Love SC weighting (for the case of SC1) in Fig.~\ref{fig:LoveSCweight_drh}.
The results are similar to the OEWG case. Deterioration of the solution is observed at larger weights of the SC, at about $\lambda=10^{5}\epsilon_\textsc{oe}^2$. The lower limit for obtaining an acceptable zero field again depends on the SNR.
Also, the GMRES stopping criterion heavily influences the SC fulfillment.  For the high-SNR/low-weighting region, the weighting needs to be relaxed to obtain a Love-current solution. In the following, we choose $\lambda=10^{3}\epsilon_\textsc{oe}^2$.

\begin{figure}[tp]
\centering
\subfloat[]{\includegraphics[width=0.32\linewidth,valign=c]{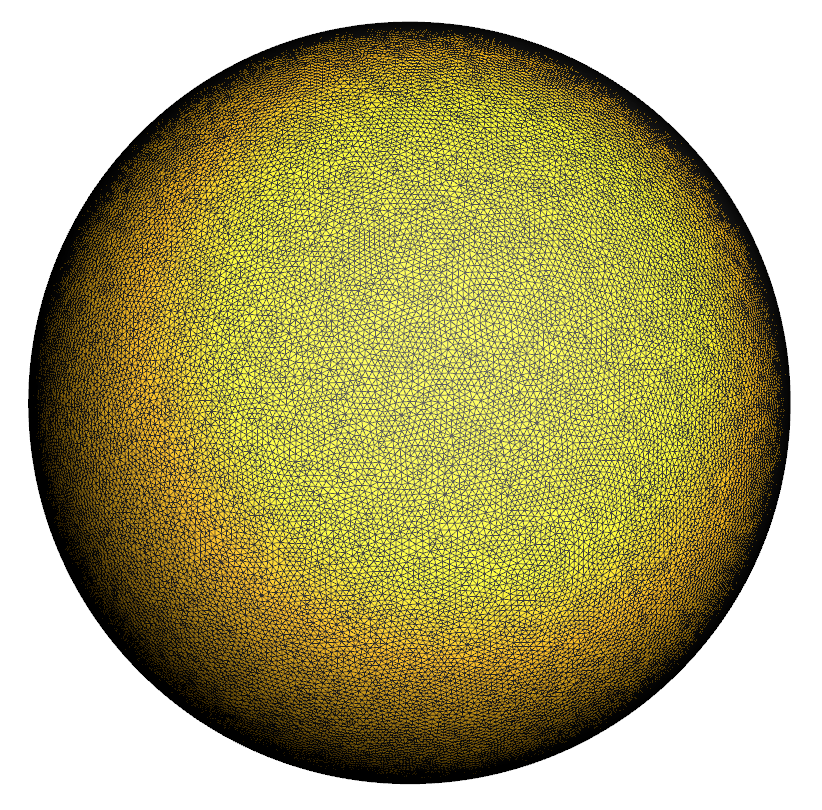}%
}\hspace*{0.225cm}
\subfloat[]{\includegraphics[width=0.27\linewidth,valign=c]{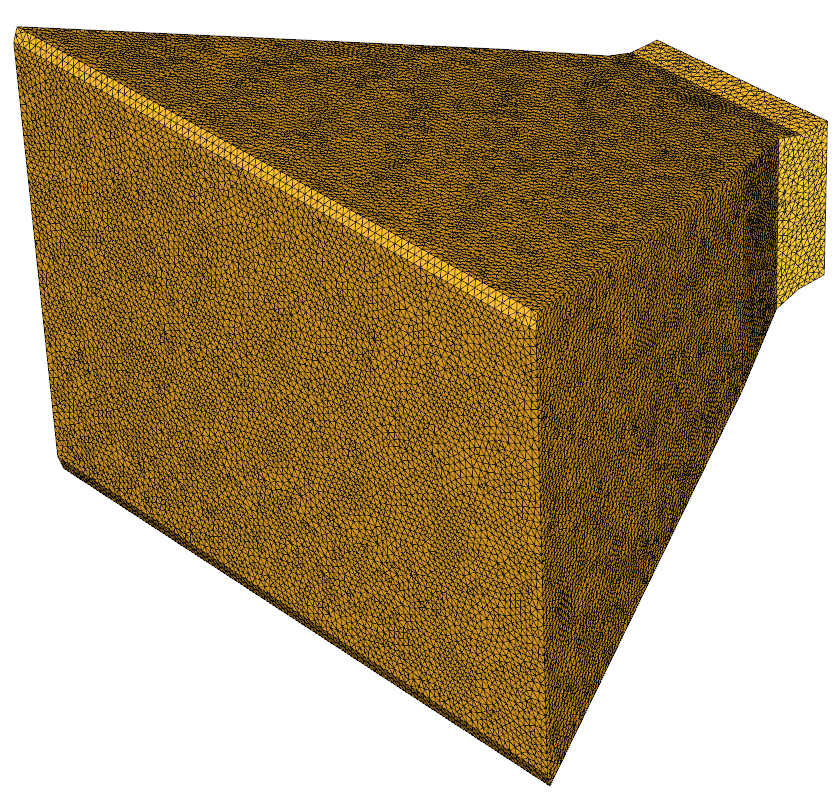}%
\vphantom{\includegraphics[width=0.32\linewidth,valign=c]{fig9a.png}}%
}\hspace*{0.225cm}
\subfloat[]{\includegraphics[width=0.27\linewidth,valign=c]{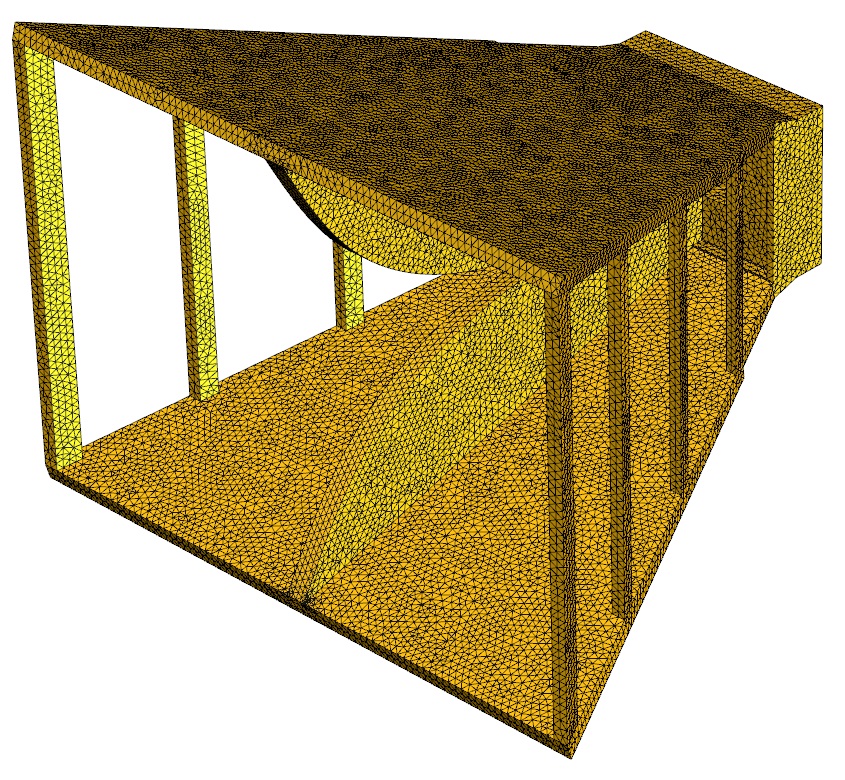}%
\vphantom{\includegraphics[width=0.32\linewidth,valign=c]{fig9a.png}}%
}
\caption{Huygens surfaces for the DRH400 simulation model: (a) a sphere, (b) a tight, close to convex hull, and (c) an exact model.
\label{fig:models}}
\end{figure}

\begin{figure}[tp]
\centering
\subfloat[]{\includegraphics[]{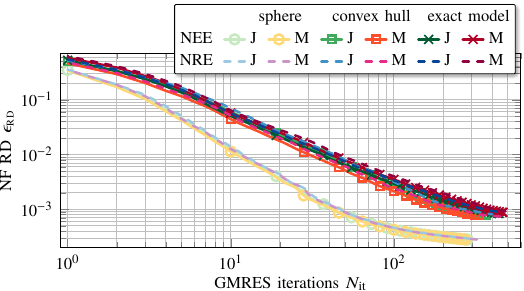}}\\
\subfloat[]{\includegraphics[]{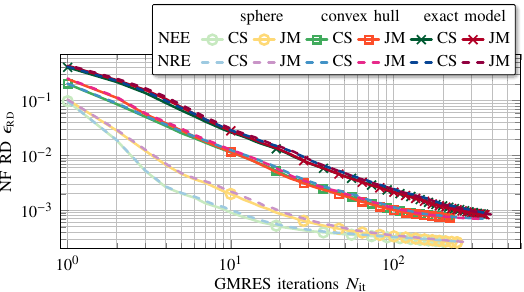}}\\
\subfloat[]{\includegraphics[]{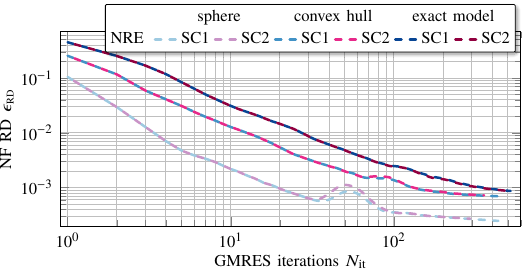}}\\
\subfloat[]{\includegraphics[]{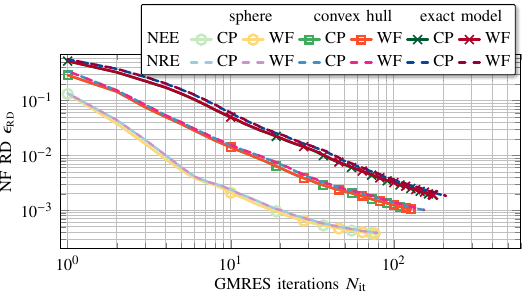}}
\caption{Iterative solver convergenve of the \ac{FIAFTA} with equivalent surface currents. (a) J and M solutions. (b) JM and CS solutions. (c) Love SCs. (d) CP and WF-CP solutions.
\label{fig:it-solv-conv}}
 \end{figure}

For further investigations, white Gaussian noise with $\epsilon_\textsc{oe}=10^{-3}$ is added to the simulated NF. 
The inverse problem is solved for three different equivalent surfaces of the \ac{AUT}: the minimum sphere, a tight hull (close to the smallest convex hull) around the \ac{AUT}, and an exact geometrical representation, see Fig.~\ref{fig:models}.
The respective number of triangles is $100\,238$, $53\,286$, and $64\,508$. 
The distance of the hull and the (scaled) exact equivalent model to the simulation model is about \SI{1}{\mm}, which is about $\lambda/50$ at the simulation frequency.
We know from the previous simulations and from~\cite{kornprobst_measurementerror_2019,Kornprobst2019Love} that the solver residual of the NRE is arbitrary. Hence, we evaluate the NF RD in each step of the iterative solver while the stopping is still based on the NRE residual. 
We consider the same surface-source types as before and a relative solver stopping  criterion, for the NRE if $\Deltaup\epsilon_\mathrm{res}>0.995$ three times in a row, and for the NEE a somewhat relaxed version with $\Deltaup\epsilon_\mathrm{res}>0.999$.  
In order to estimate the accuracy and conditioning of the source representations, the iterative solution process is studied in Fig.~\ref{fig:it-solv-conv}.  
A summary of results is given in Tab.~\ref{tab:side-weight-DRH} for the convex hull model.
\begin{table}[tp]
\caption{Source type comparison on a convex hull around the AUT, $\epsilon_\textsc{oe}=10^{-3}$, for the (\textup{a})~NRE with $\Deltaup\epsilon_\mathrm{res}=0.995$ and (\textup{b})~the NEE with $\Deltaup\epsilon_\mathrm{res}=0.999$.\label{tab:side-weight-DRH}}
\centering
\subfloat[]{
\begin{tabular}{L{17.5mm} R{4mm}R{10.5mm}R{9mm}R{10.5mm}R{10.5mm}}
\toprule[1pt]
source type&   $N_\mathrm{it}$ &  $\epsilon_\mathrm{res}$ &$\epsilon_\textsc{rd}$/$\epsilon_\textsc{oe}$ & $\epsilon_\mathrm{\textsc{zf},avg}$ & $\epsilon_\mathrm{\textsc{ff},max}$ \\
\cmidrule[0.5pt](lr){1-1} \cmidrule[0.5pt](lr){2-2}  \cmidrule[0.5pt](lr){3-3}  \cmidrule[0.5pt](lr){4-4} \cmidrule[0.5pt](lr){5-5} \cmidrule[0.5pt](lr){6-6}
J -- NRE         &\vb  $402$  &\nt $2.6\cdot10^{-6}$  &\gd  $0.81$  &\vb   $ 7.0$\,dB    &\vg  $-70.8$\,dB\\[0.2ex]
M -- NRE         &\vb  $451$  &\nt $1.6\cdot10^{-6}$  &\bd  $0.74$  &\vb   $ 6.7$\,dB    &\vg  $-72.0$\,dB  \\[0.2ex]
JM -- NRE        &\bd  $336$  &\nt $1.2\cdot10^{-6}$  &\bd  $0.70$  &\bd   $ 0.0$\,dB    &\vg  $-73.1$\,dB  \\[0.2ex]
CS -- NRE        &\bd  $336$  &\nt $1.2\cdot10^{-6}$  &\bd  $0.71$  &\gd  $ -6.8$\,dB    &\vg  $-72.5$\,dB  \\[0.2ex]
CP -- NRE        &\vg  $157$  &\nt $1.2\cdot10^{-5}$  &\vg  $0.99$  &\vg  $-27.4$\,dB    &\gd  $-66.5$\,dB  \\[0.2ex]
WF-CP -- NRE     &\vg  $153$  &\nt $1.3\cdot10^{-5}$  &\gd  $1.02$  &\vg  $-27.2$\,dB    &\gd  $-66.1$\,dB  \\[0.2ex]
SC1 -- NRE       &\vb  $451$  &\nt $1.7\cdot10^{-5}$  &\bd  $0.69$  &\vg  $-30.0$\,dB    &\vg  $-73.0$\,dB  \\[0.2ex]
SC2   -- NRE     &\bd  $357$  &\nt $3.3\cdot10^{-6}$  &\bd  $0.72$  &\gd  $ -9.4$\,dB    &\vg  $-71.7$\,dB  \\
\bottomrule[1pt]
\end{tabular}
}\\
\subfloat[]{
\begin{tabular}{L{17.5mm} R{4mm}R{10.5mm}R{9mm}R{10.5mm}R{10.5mm}}
\toprule[1pt]
J -- NEE         &\bd  $368$  &\bd  $7.4\cdot10^{-4}$  &\bd  $0.74$  &\vb    $6.8$\,dB    &\vg  $-71.1$\,dB\\[0.2ex]
M -- NEE         &\bd  $330$  &\gd  $8.0\cdot10^{-4}$  &\gd  $0.80$  &\vb    $6.7$\,dB    &\vg  $-69.0$\,dB  \\[0.2ex]
JM -- NEE        &\gd  $222$  &\bd  $7.3\cdot10^{-4}$  &\bd  $0.73$  &\bd    $0.0$\,dB    &\vg  $-73.3$\,dB  \\[0.2ex]
CS -- NEE        &\gd  $228$  &\bd  $7.5\cdot10^{-4}$  &\bd  $0.75$  &\gd  $ -6.8$\,dB    &\vg  $-71.5$\,dB  \\[0.2ex]
CP -- NEE        &\vg  $127$  &\gd  $1.0\cdot10^{-3}$  &\gd  $1.04$  &\vg  $-28.1$\,dB    &\gd  $-66.4$\,dB  \\[0.2ex]
WF-CP -- NEE     &\vg  $128$  &\gd  $1.0\cdot10^{-3}$  &\gd  $1.05$  &\vg  $-27.8$\,dB    &\gd  $-66.6$\,dB  \\
\bottomrule[1pt]
\end{tabular}
}\vspace*{-0.5cm}
\end{table}%

The number of solver iterations and the convergence rate give some insight into the conditioning. The NF RD and FF error provide insight  into the achievable accuracy levels. 
The largest differences in the solution behavior are observed between the three choices of the reconstruction surface: The sphere leads to the fastest convergence (i.e., the best conditioning), the convex hull performs worse for any source type, and the exact geometrical representation is even slower.
Investigations on the shape of the reconstruction surface have already been performed in~\cite{quijano20103d,leone2018application,knapp2019near,leone2020inverse}, in part for echo suppression applications. The shape of the equivalent source surface may introduce weakly radiating currents (evanescent modes) if the surface is non-convex. Hence, the reconstruction surface has to be chosen accordingly. In the comparison at hand including various equivalent source types, the influence of the Huygens surface shape is certainly also an interesting factor of influence.
Interestingly, the convergence curves of Love-current representations and pure electric-current solutions are very similar if the reconstruction surface comes close to the conducting antenna model. 
From a physical point of view, this kind of similar behavior is expected due to the fact that both systems of equations yield the same solution of purely electric currents for the same right-hand side, hence the matrices have to be very similar.

Looking at the accuracy of the equivalent surface choices, we can state that the sphere leads to overfitting (NF RD below the noise level of $10^{-3}$, see Fig.~\ref{fig:it-solv-conv}) and, hence, reduced accuracy. 
The FF errors of the sphere model are in the range of $-44$\,dB to $-54$\,dB\,---\,the worst results are obtained for J and M solutions. 
The convex hull offers a RD of about $10^{-3}$ and FF errors of mostly below $-70$\,dB, which is a significant improvement  (\emph{processing gain}).
The exact model is able to provide better diagnostic information, but requires a lot of detailed information about the AUT. The optimal reconstruction deviation of $10^{-3}$ is reached, see Fig.~\ref{fig:it-solv-conv}, and the FF errors are about 1\,dB to 2\,dB worse than for the convex hull\,---\,i.e., absolutely comparable.
Overall, for best accuracy and reasonable conditioning (fast iterative solver convergence), a convex hull is the reasonable choice.

\section{Source Reconstruction for Measurement Data of a Reflector Antenna}

\begin{figure}[tp]
\centering
\subfloat[]{\includegraphics[height=0.35\linewidth,valign=c]{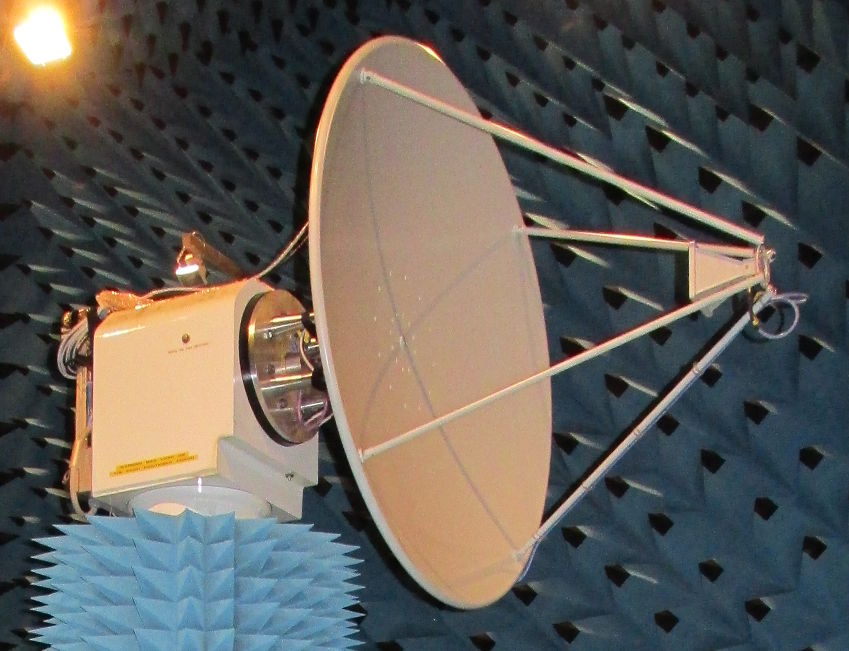}%
}\hspace*{0.45cm}
\subfloat[]{\includegraphics[height=0.34\linewidth,valign=c]{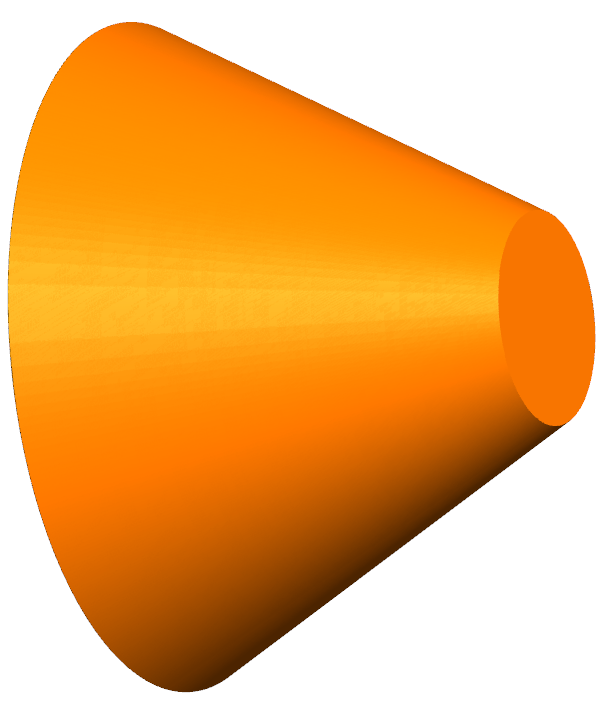}%
\vphantom{\includegraphics[height=0.35\linewidth,valign=c]{fig11a.jpg}}%
}
\caption{Measured reflector antenna (a) inside the anechoic chamber and (b) as a reconstruction model.
\label{fig:refl}}
\end{figure}

Spherical \ac{NF} measurements of a parabolic reflector antenna have been conducted at \SI{18}{GHz} with a DRH18 probe in the measurement facilities of Rohde \& Schwarz~\cite{Neitz2017,RSRefl,DRH18}. 
The \ac{AUT} exhibits a diameter of \SI{1.23}{m} and the measurement distance was \SI{5}{m}. 
Based on our previous insights, only a convex hull is considered as reconstruction surface. 
The AUT inside the anechoic chamber and the reconstruction surface are shown in Fig.~\ref{fig:refl}.

The number of NF measurement samples is $812\,702$ (for two orthogonal polarizations), which also equals the number of unknowns for all NEE variants. 
For the NRE, the D-SH formulation has $14\,143\,275$ unknowns. 
The employed mesh comprises $4\,289\,139$ RWG unknowns, with double the number of unknowns for the JM approach.

\begin{figure}[tp]
\centering
\subfloat[]{\includegraphics[]{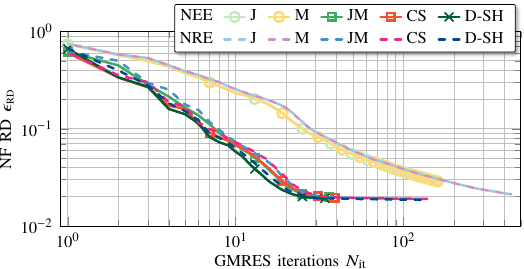}}\\
\subfloat[]{\includegraphics[]{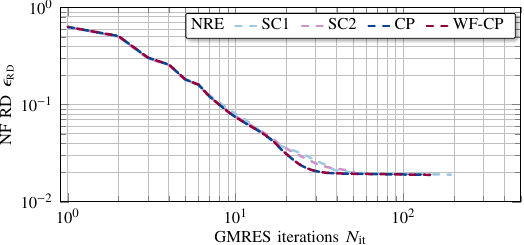}}\\
\subfloat[]{\includegraphics[]{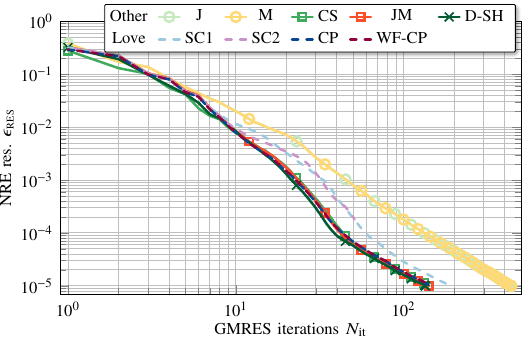}}\\
\caption{Iterative solver convergence for the source reconstruction of reflector antenna NF measurements. (a)~NF RD of non-physical surface source representations, for both NRE and NEE. (b)~NF RD of Love-current formulations, for the NRE. (c)~Solver residual of the NRE solvers.
\label{fig:refl_conv}}
\end{figure}

An iterative solver convergence study is shown in Fig.~\ref{fig:refl_conv}.
The stopping criterion was chosen relative with a residual decrease slower than $0.997$ three times in a row. 
Additionally, the NRE solver was stopped once a residual threshold of $10^{-5}$ was reached.
The RD estimates the OE to about $2$\si{\percent} consistently among all solvers.
All formulations show a comparable convergence behavior with a slight advantage for the NEE. 
The control of the solver termination, once a stagnating NF RD is reached, works much better for the NEE.
For instance, the JM-NEE version stops at an NF RD of $1.96$\si{\percent} after $36$ iterations, while the JM-NRE version stops after $141$ iterations at a residual of $9.9\cdot10^{-6}$ and an NF RD of $1.89$\si{\percent}. The NRE reaches the same NF RD as the NEE with convergence at 39 iterations. 
The JM, CS, and D-SH variants perform very similarly. The J and M versions show a worse convergence.

The Love-current SC formulations show a worse convergence than the JM, CS and D-SH formulations. 
The CP and WF-CP variants show a comparable convergence.
However, it has to be kept in mind that each iteration is computationally more costly.

For comparison, a fully probe-corrected spherical transformation~\cite{Mauermayer2018Spherical} with the NRE takes $35$ iterations for a residual of $10^{-4}$ and stops at an NF RD of $1.67$\si{\percent}, i.e., it suffers a bit from overfitting due to the inherently limited source localization.

\begin{figure}[tp]
\centering
\includegraphics[]{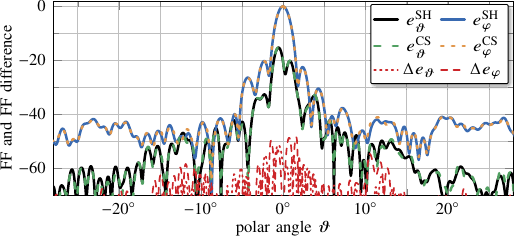}
\caption{Transformed FFs of a SH expansion solved with the NRE and a CS surface-source solution of the NEE. 
\label{fig:ff}}
\end{figure}

Transformed FFs of the spherical transformation and the NEE-CS source reconstruction are shown in Fig.~\ref{fig:ff}, with the relative magnitude deviation reaching up to $-48.5$\,dB. 
The deviation between the various surface-source formulations is below this level.
Hence, an accuracy analysis for measurement data is not feasible.
We can, however, compare some of the deviations.
The maximum deviation between the CS solutions with the NEE and the NRE is at $-59.0$\,dB, between the JM and the CS solutions with the NEE at $-72.1$\,dB, between the NRE-J and the NEE-CS solutions is at $-53.8$\,dB, between the NRE-CP and the NEE-CS solutions at $-59.0$\,dB, and between the NRE-SC1 and the NEE-CS solutions at $-59.9$\,dB.
All these values are below the measurement accuracy.

\section{Conclusion}
We have discussed and analyzed the inverse surface-source problem related to near-field antenna measurements in detail, with focus on various source representations.

Regarding the accuracy of the reconstructed fields, all source types performed in a very similar way. The versions with zero field SC give good results if the SC weighting is chosen suitably. 
The accuracy is also heavily influenced by the chosen reconstruction surface, where a convex hull around the AUT provided the best source localization and solution accuracy.

The conditioning of the surface-source problem mostly depends on the choice of the reconstruction surface. 
Larger convex surfaces go hand in hand with a better conditioning. However, this inevitably leads to imperfect reconstructions due to the fact that a larger portion of the OE can be mapped to a wrong part in the solution.
The various source representations only have a minor influence on the conditioning; however, pure electric or magnetic current solutions are inferior. In the case of placing the equivalent source surface onto a conducting part of the AUT, the Love-current solutions show the same worsened conditioning as the pure electric current solution.

The inverse problem was solved iteratively in a least-squares sense by utilizing normal equations. 
While the wide-spread NRE may suffer from overfitting, the Love SC can help to prevent this by introducing a stopping threshold.
Yet, this is achieved with much lower computational effort by employing the NEE version with any kind of equivalent sources.

While Love currents offer additional diagnostic capabilities, we could not identify benefits during the solution process.
Hence, post processing techniques for the field visualization on the AUT surface seem to be the more viable method.

\appendix

\section{Summary of Abbreviations}
The abbreviations employed throughout the paper are given in Table~\ref{tab:acs}.
\begin{table}%
\caption{Summary of Abbreviations. (\textup{a})~Inverse problem notation. (\textup{b})~Equivalent source type notation.\label{tab:acs}}%

\centering
\subfloat[]{%
\begin{tabular}{llr}
\toprule[1pt]
Abbr.\!\!&  long version & explanation \\
\cmidrule[0.5pt](lr){1-1} \cmidrule[0.5pt](lr){2-2} \cmidrule[0.5pt](lr){3-3} 
NE     & normal system of equations      &   either NRE or NEE \\[0.2ex]
NRE  \!& normal-residual system of eqs.  &  overdetermined case \\
       &                                 &    or Tikhonov regularization  \\[0.2ex]
NEE  \!& normal-error system of eqs.     &  underdetermined case \\[.2ex]
OE     & observation error               &  meas.\ errors, noise, etc.  \\[0.2ex]
RD     & reconstruction deviation        &  difference to measured NFs\\ 
\bottomrule[1pt]
\end{tabular}%
}\\
\subfloat[]{%
\begin{tabular}{ll}
\toprule[1pt]
Abbr.&  long version  \\
\cmidrule[0.5pt](lr){1-1} \cmidrule[0.5pt](lr){2-2} 
J      & purely electric currents                          \\[0.2ex]
M      & purely magnetic currents                          \\[0.2ex]
JM     & unconstrained electric and magnetic currents                          \\[0.2ex]
CS     & weak-form combined sources       \\[0.2ex]
SC1    & Love-current side constraint, MFIE-alike    \\[0.2ex]
SC2    & Love-current side constraint, EFIE-alike    \\[0.2ex]
SC3    & Love-current side constraint, CFIE-alike    \\[0.2ex]
CP     & Calderón projector                \\[0.2ex]
WF-CP  & improved Calderón projector       \\[0.2ex]
SH     & spherical-harmonics expansion     \\[.2ex]
D-SH   & distributed SHs with MLFMM        \\
\bottomrule[1pt]
\end{tabular}%
}
\end{table}%
\bibliographystyle{IEEEtran}

\bibliography{IEEEabrv,ref}

\begin{IEEEbiography}
 [{\includegraphics[width=1in,height=1.25in,clip,keepaspectratio]{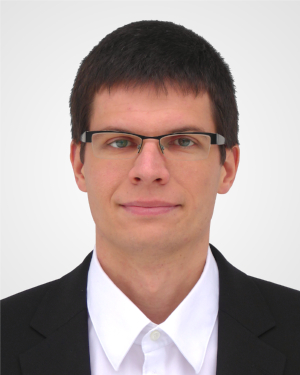}}]
{Jonas Kornprobst} (Graduate Student Member, IEEE) received the B.Eng. degree in electrical engineering and information technology from the University of Applied Sciences Rosenheim, Rosenheim, Germany, in 2014, and the M.Sc. degree in electrical engineering and information technology from the Technical University of	Munich, Munich, Germany, in 2016. 

Since 2016, he has been a Research Assistant with the Chair of High-Frequency Engineering, Department of Electrical and Computer Engineering, Technical University of Munich. 
His current research interests include numerical electromagnetics, in particular integral equation	methods, antenna measurement techniques, antenna and antenna array design, as well as microwave circuits.
\end{IEEEbiography}
\begin{IEEEbiography}
 [{\includegraphics[width=1in,height=1.25in,clip,keepaspectratio]{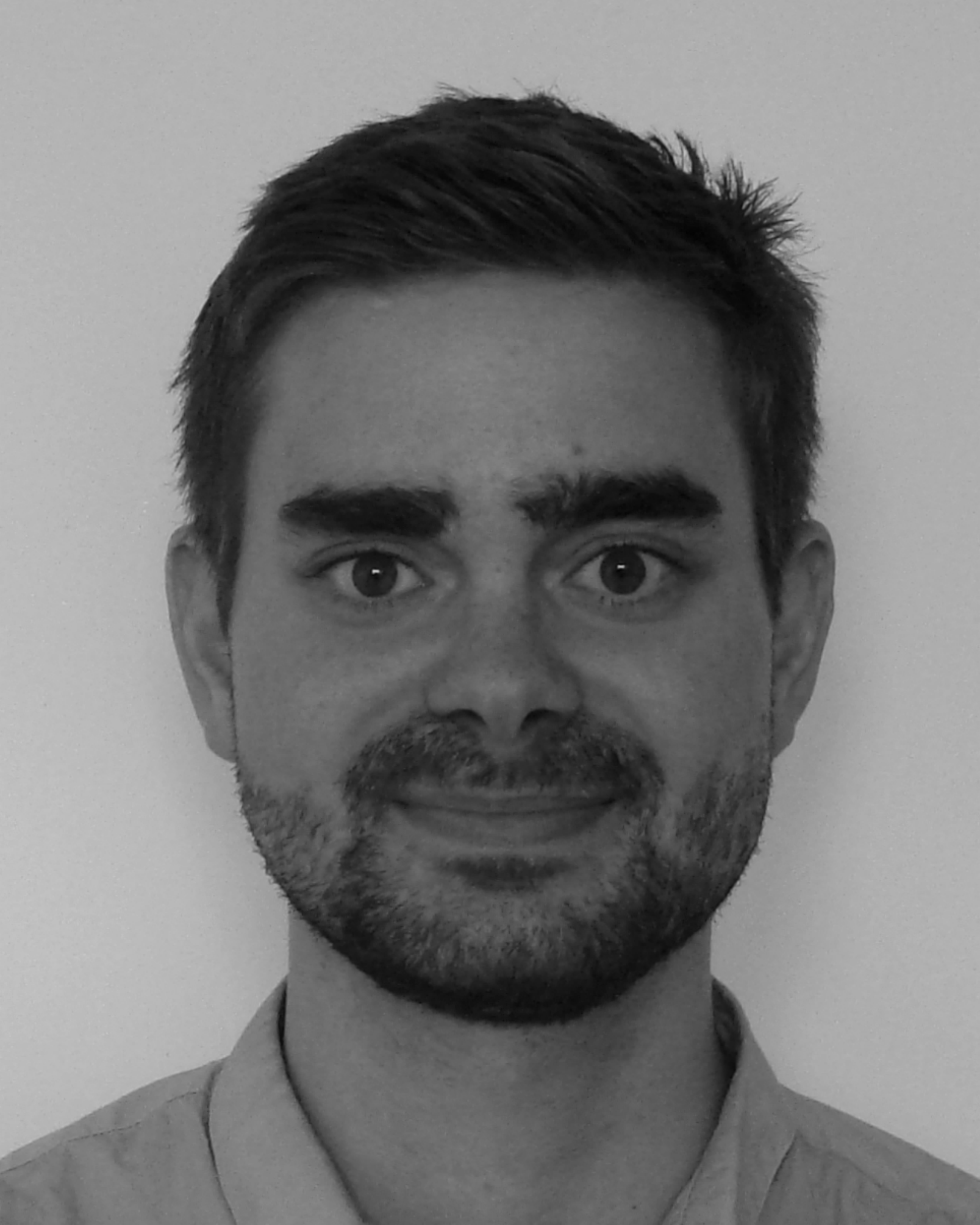}}]
{Josef Knapp} (Graduate Student Member, IEEE) received the M.Sc. degree in electrical engineering and information technology from the Technical University of Munich, Munich, Germany, in 2016. 

Since 2016, he has been a Research Assistant at the Chair of High-Frequency Engineering, Department of Electrical and Computer
Engineering, Technical University of Munich. 
His research interests include inverse electromagnetic problems, computational electromagnetics, antenna measurement techniques in unusual environments, and field transformation techniques.
\end{IEEEbiography}
\begin{IEEEbiography}
 [{\includegraphics[width=1in,height=1.25in,clip,keepaspectratio]{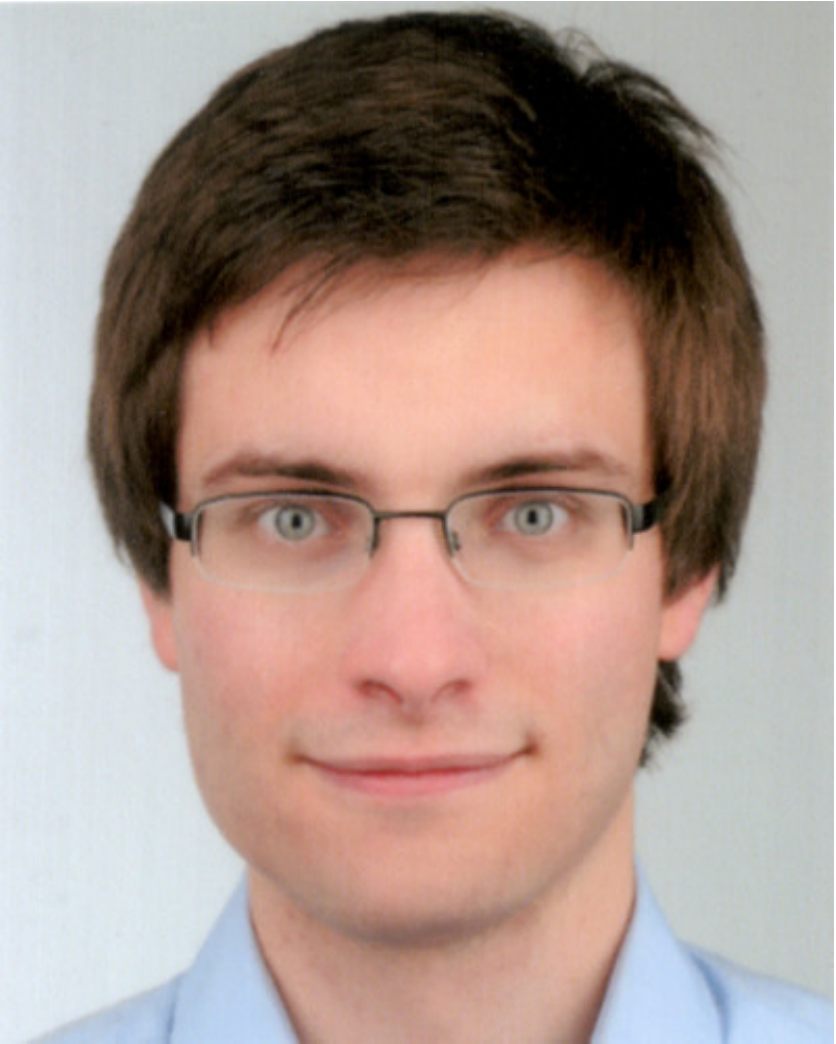}}]
{Raimund A. M. Mauermayer} (Graduate Student Member, IEEE) received the Dipl.-Ing. degree in electrical engineering and information technology from the Technical University of Munich, Munich, Germany, in 2012.

From 2012 to 2018, he was a Research Assistant with the Chair of High-Frequency Engineering, Technical University of Munich. 
In 2019, he joined the Department of Antenna Design and Measurement, Mercedes-Benz AG, Sindelfingen, Germany. 
His research interests are near-field far-field transformation and antenna measurement techniques.
\end{IEEEbiography}

\vspace*{5cm}

\begin{IEEEbiography}
 [{\includegraphics[width=1in,height=1.25in,clip,keepaspectratio]{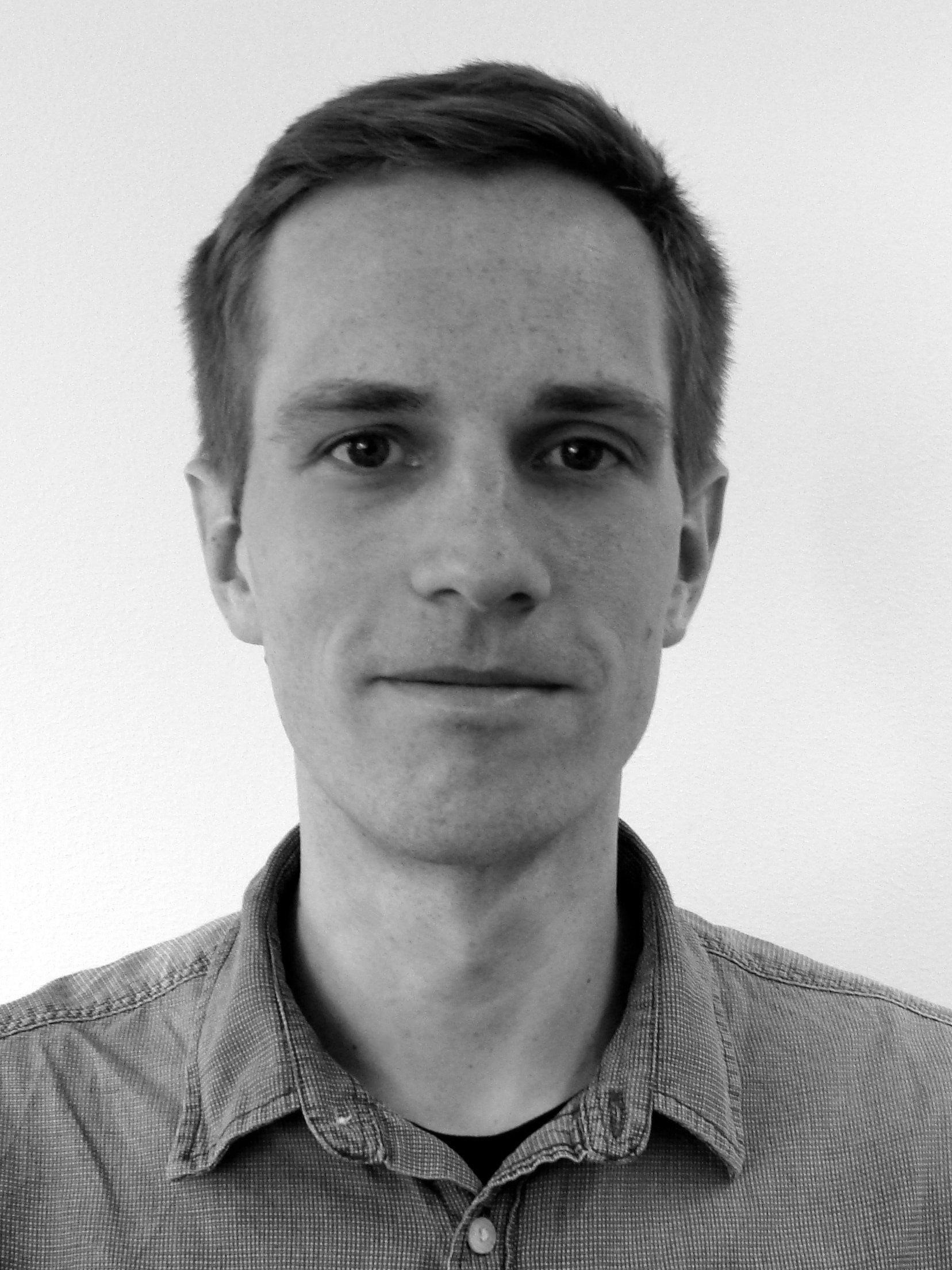}}]
{Ole Neitz} (Graduate Student Member, IEEE)  received  the  Dipl.-Ing.  degree  in electrical  engineering  and  information  technology from  the  University  of  Bremen,  Bremen,  Germany in 2013 and the Dr.-Ing. degree from the Technical University of Munich, Munich, Germany, in 2020. 

From 2013 to 2019, he was a Research Assistant at  the  Chair  of  High-Frequency  Engineering,  Department  of  Electrical  and  Computer  Engineering, Technical University of Munich. 
In 2019, he joined the  Antenna  Development  Department  of  Rohde \& Schwarz  GmbH  \&  Co.  KG,  Munich.  
His  current research  interests  include  antenna  development  and measurement  techniques,  as  well  as  near-field  scattering  and  imaging  problems.
\end{IEEEbiography}
\begin{IEEEbiography}
 [{\includegraphics[width=1in,height=1.25in,clip,keepaspectratio]{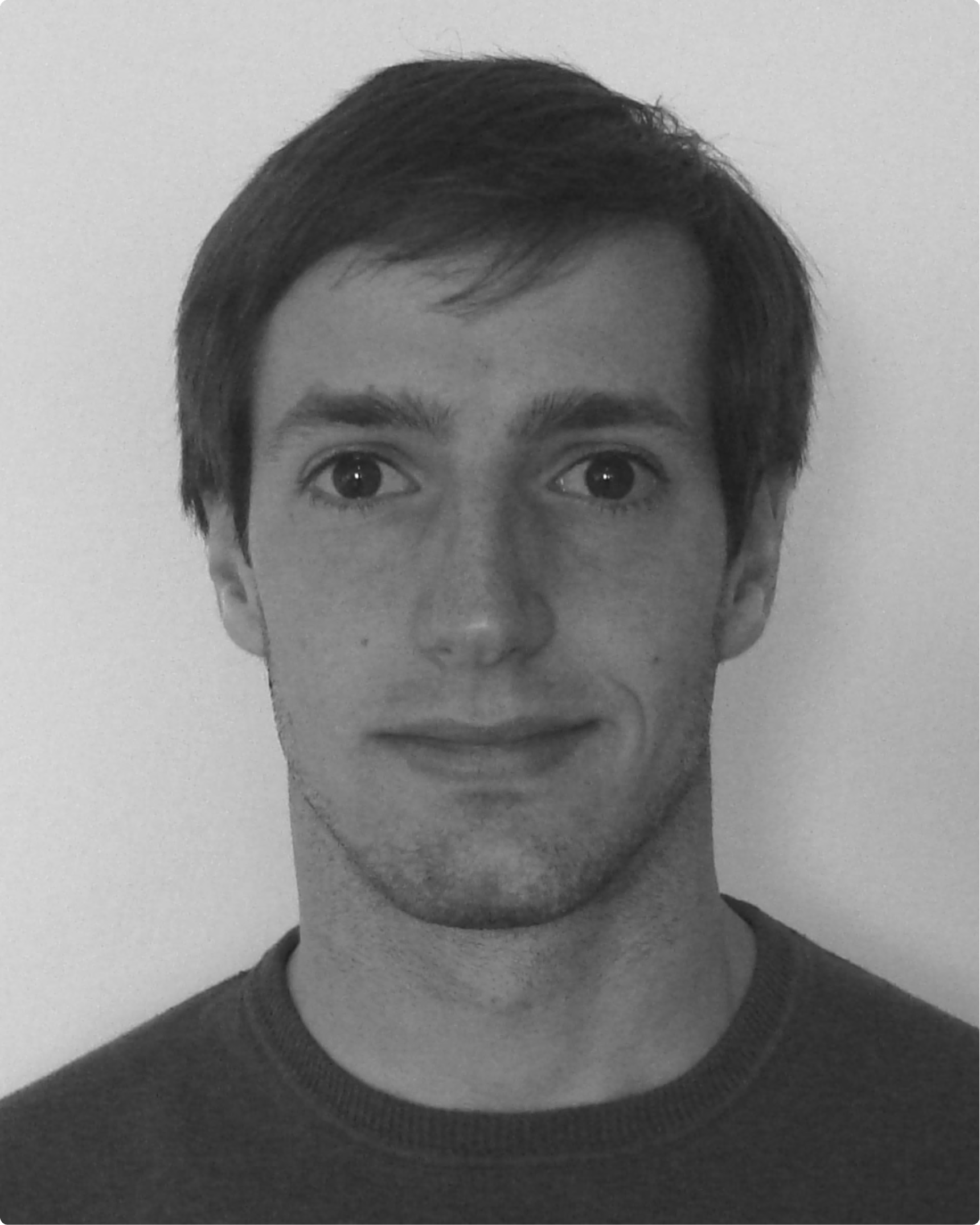}}]
{Alexander Paulus} (Graduate Student Member, IEEE) received the M.Sc. degree in electrical engineering and information technology from the Technical University of Munich, Munich, Germany, in 2015. 

Since 2015, he has been a Research Assistant at the Chair of High-Frequency Engineering, Department of Electrical and Computer
Engineering, Technical University of Munich. 
His research interests include inverse electromagnetic problems, computational electromagnetics and antenna measurement techniques.
\end{IEEEbiography}
\begin{IEEEbiography}
 [{\includegraphics[width=1in,height=1.25in,clip,keepaspectratio]{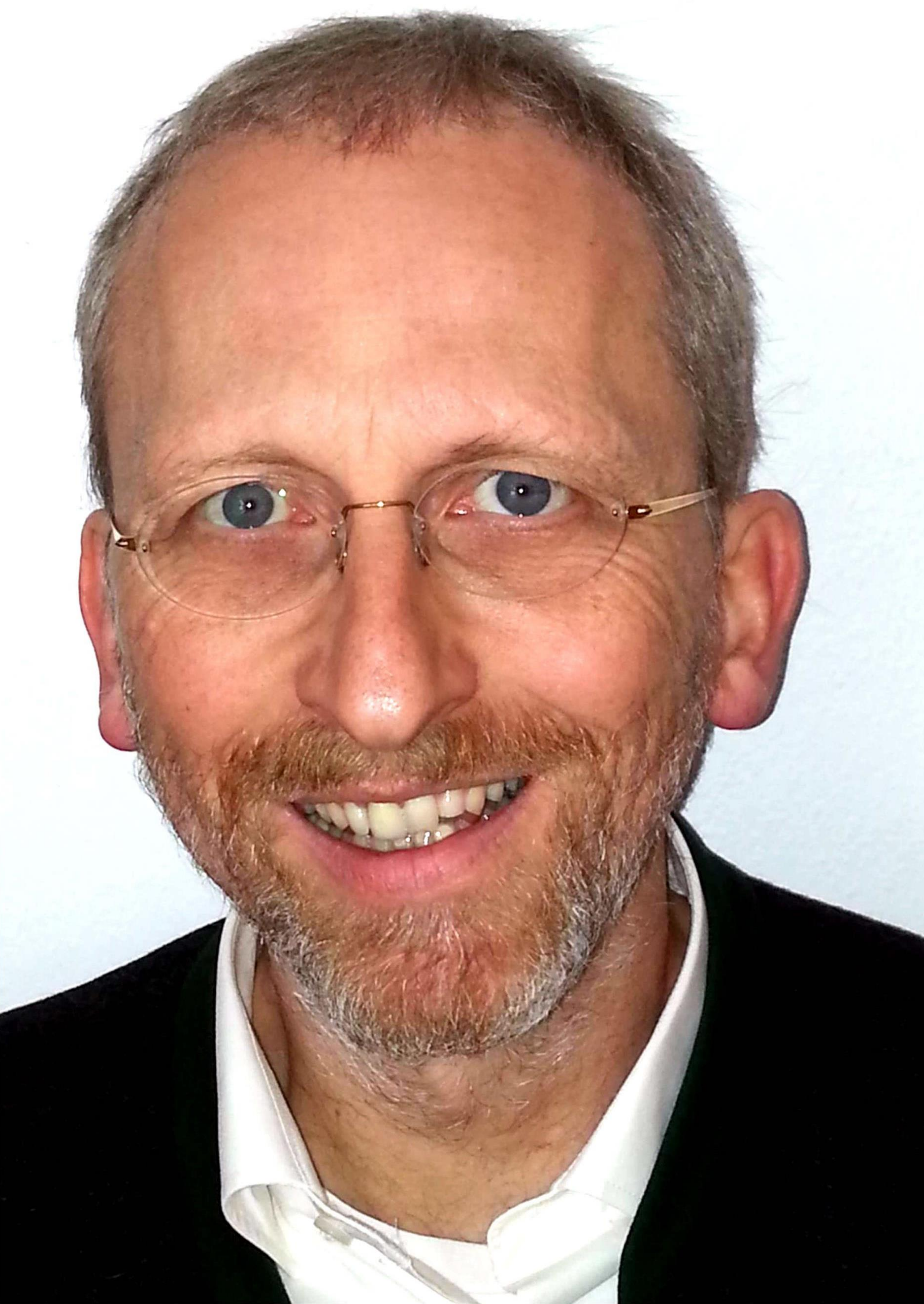}}]
{Thomas F.~Eibert} (Senior Member, IEEE) received the Dipl.-Ing.\,\,(FH) degree from Fachhochschule N\"urnberg, Nuremberg, Germany, the Dipl.-Ing.~degree from Ruhr-Universit\"at Bochum, Bochum, Germany, and the Dr.-Ing.~degree from Bergische Universit\"at Wuppertal, Wuppertal, Germany, in 1989, 1992, and 1997, all in electrical engineering. 

He is currently a Full Professor of high-frequency engineering at the Technical University of Munich, Munich, Germany. 
His current research interests include numerical electromagnetics, wave propagation, measurement and field transformation techniques for antennas and scattering as well as all kinds of antenna and microwave circuit technologies for sensors and communications.		
\end{IEEEbiography}

\vspace*{5cm}
\end{document}